\documentclass[prb,aps,twocolumn,superscriptaddress,nofootinbib]{revtex4}

\usepackage{epsfig}
\usepackage{amsfonts}
\usepackage{latexsym}
\usepackage{amsmath}
\usepackage{amssymb}
\usepackage{slashed}
\usepackage{longtable}
\usepackage{lipsum}
\usepackage{multirow}
\usepackage{array}

\usepackage{graphicx}% Include figure files
\usepackage{hyperref}
\usepackage{pstool}

\numberwithin{equation}{section}

 \def\p{\partial}

\newcommand{\bea}{\begin{eqnarray}}
\newcommand{\eea}{\end{eqnarray}}
\newcommand{\be}{\begin{equation}}
\newcommand{\ee}{\end{equation}}
\newcommand{\ba}{\begin{align}}
\newcommand{\ea}{\end{align}}

\def\Or[#1]{{\text{O}}\left({#1}\right)}
\def\dotl[#1,#2]{\left\langle #1, #2 \right\rangle}
\def\dotlb[#1,#2]{[ #1, #2 ]}
\def\dotp[#1,#2]{(#1) \cdot (#2)}
\def\aff[#1,#2]{\hat{#1}(#2)}
\def\n4sym{{\cal N}=4 SYM}
\def\>{\rangle}
\def\<{\langle}
\def\weight[#1,#2,#3]{\{(#1),#2,#3\}}
\def\ads[#1]{$\text{AdS}_{#1}$}

  \makeatletter
  \let\over=\@@over \let\overwithdelims=\@@overwithdelims
  \let\atop=\@@atop \let\atopwithdelims=\@@atopwithdelims
  \let\above=\@@above \let\abovewithdelims=\@@abovewithdelims
\usepackage{xcolor}

\begin{document}

\preprint{}

\title{Boundary condition and geometry engineering in electronic hydrodynamics}

\author{Roderich Moessner}
\affiliation{Max-Planck-Institut  f\"ur Physik komplexer Systeme, N\"othnitzer Str. 38, 01187 Dresden, Germany}
\author{Nicol\'as Morales--Dur\'an} 
 \affiliation{Department of Physics, The University of Texas at Austin, Austin, TX 78712, USA}
\author{Piotr Sur\'owka}
\affiliation{Max-Planck-Institut  f\"ur Physik komplexer Systeme, N\"othnitzer Str. 38, 01187 Dresden, Germany}
\author{Piotr Witkowski} 
\affiliation{Max-Planck-Institut  f\"ur Physik komplexer Systeme, N\"othnitzer Str. 38, 01187 Dresden, Germany}

\begin{abstract}
We analyze the role of boundary geometry in
viscous electronic hydrodynamics. We address the twin questions of how
boundary geometry impacts flow profiles, and how one can engineer
boundary conditions -- in particular the effective slip parameter --
to manipulate the flow in a controlled way.  We first propose a
micropatterned geometry involving finned barriers, for which we show
by an explicit solution that one can obtain effectively no-slip
boundary conditions regardless of the detailed microscopic nature of
the channel surface.  Next we analyse the role of mesoscopic boundary
curvature on the effective slip length, in particular its impact on
the Gurzhi effect.  Finally we investigate a hydrodynamic flow through a
circular junction, providing a solution, which suggests an
experimental set-up for determining the slip parameter. We find that its
transport properties differ qualitatively from the case of
ballistic conduction, and thus presents a promising setting for
distinguishing the two. 
\end{abstract}

\maketitle
\section{Introduction}
The field of electronic transport phenomena has been greatly enriched
by the idea that for sufficiently strong electron-electron scattering,
a description in terms of an effectively viscous hydrodynamics becomes
appropriate. This studies the flow of conserved quantities such as
mass, charge or energy (heat).  Proposed a long time ago by Gurzhi in
\cite{Gurzhi1963,Gurzhi1968}, only recently a family of samples clean
enough to observe a wide range of hydrodynamic effects (such as
negative local resistance, superballistic flow or a modification of the
Hall effect
\cite{Molenkamp1994,deJong1995,Crossno2016,Bandurin2016,Moll2016,KrishnaKumar2017,Guo2017,Berdyugin})
has been subjected to a systematic study. For a review of the
field of viscous electronics see \cite{Lucas2018}.

One of the important aspects of hydrodynamics as an effective
transport theory is its inherently mesoscopic character due to the
fact that the solutions of the transport equations strongly depend on
the boundary conditions. The so-called Maxwell boundary conditions, a
one parameter family of consistent boundary conditions for
hydrodynamics, read
\begin{align}
 \label{eq:GeneralBCflat}
    \left.u^t_i\right|_B=\left.\xi\,n_j\,\frac{\partial u^t_i}{\partial x_j} \right|_B \ .
\end{align}
The parameter $\xi$ is called the \emph{slip length}. This boundary condition involves the
tangent velocity $u^t$ at the boundary of flow domain and its normal
(with respect to \emph{inward} pointing vector $n_j$) derivative.

In many every-day uses of hydrodynamics, the slip
length is negligibly small, encoded by the no-slip boundary condition
that prohibits the fluid from having any tangent velocity at the
domain's boundary. However, there are situations, like in liquid
helium or microfluidics\cite{PanzerPRL,Lauga2007}, where a nonzero
slip length cannot be neglected.  

Recent experiments failing to observe the Gurzhi effect in a long
graphene channel \cite{Bandurin2016}, as well as theoretical insights
on the dependence of slip length $\xi$ on temperature
\cite{Kiselev2019}, suggest that a similar situation may exist for
viscous electronics. In that case the question of determining and
correctly treating the boundary condition becomes crucial for both
further theoretical developments and possible practical applications
of the field.

Whereas the microscopic\footnote{To avoid confusion
  with length scale description, let us stress here that when we write
  about \emph{microscopic} effects, we mean atomic-scale effects;
  \emph{mesoscopic} will denote scales around the micrometer scale,
  and by \emph{macroscopic} we mean things measurable at everyday
  length-scales, e.g.\ temperature. We also sometimes use the word
  \emph{sub-mesoscopic} to denote intermediate scales between micro-
  and mesoscopic, i.e. larger than atomic, but smaller than, e.g.,
  typical sample dimensions.} slip in viscous electronic systems has been
investigated in detail \cite{Kiselev2019}, there is another aspect of
the boundary condition \ref{eq:GeneralBCflat} associated with the
{\it geometry} of the boundary, studied by Einzel, Panzer and Liu
\cite{PanzerPRL}. If the boundary of a channel is not flat but has
some curvature, it modifies the boundary condition by replacing the
microscopic slip length $\xi$ by an effective parameter
$\xi_{\text{eff}}$ that is a function of the local curvature. For a
mesoscopic sample the curvature can be either mesoscopic (i.e. of
order of characteristic system size) or sub-mesoscopic (i.e. much
smaller than system size yet bigger than momentum conserving
scattering mean free path). To our knowledge, this aspect of boundary
conditions has not been analysed in the context of viscous
electronics.

In this work we address a group of issues concerning boundary
conditions of viscous electronic flow relating to their nature,
observability, and tunability. We start in section
\ref{sec:engineering} with answering a practical question: can one,
independently of the nature of microscopic boundary condition, perform
some micro-structuring of the boundary that would effectively yield a
well controlled boundary condition? The answer to that question turns
out to be affirmative as a relatively simple boundary patterning turns
out to mimic the classical no-slip boundary condition.  Then, in section
\ref{sec:curvedbdr} we discuss  the EPL
(Einzel-Panzer-Liu) boundary condition. We present modifications of
the basic (Hagen-)Poiseuille flow to account for effective geometric
slips on the boundaries. The corresponding conductance for a couple of
test parameters is presented. Since the results obtained in that
section indicate a possible breakdown of the theory, we perform also a
linear stability analysis of those viscous flow solutions which, as technically involved, is relegated to Appendix \ref{sec:stability}.  

The next
section, \ref{sec:junctions}, is devoted to studying the effects of nonzero
slip in a flow through a circular junction -- a set-up that was
previously investigated in the framework of ballistic transport in
semiconductors\cite{Marcus1992,Ishio1995,Schwieters1996}. We present
qualitative differences between ballistic and hydrodynamic transport
in such a set-up and propose a measurement protocol that allows one to
directly access the slip length. The local conductivity is a
non-monotonic function of that parameter and, in turn, also 
of temperature. 
We close the main text with
conclusions and a discussion. Due to multiplicity of techniques used in
this work we supplement the text with a couple appendices in which we
elaborate on the technical side of our computations.

\section{Boundary condition engineering}\label{sec:engineering}
Recent theoretical analysis \cite{Kiselev2019} suggests that the microscopic slip length exhibits strong temperature dependence, and is divergent when $T\rightarrow{}0$. This result, implying that for low temperatures slip length can be of order of the sample size, is backed up by some experimental data \cite{Bandurin2016}. 

One can ask: why is 	the issue of boundary conditions so important? The answer stems from the fact that the viscous force is proportional to the \emph{gradient} of velocity so, in any set-up where a flow is locally parallel to the boundary, any nonzero slip length will substantially reduce the local resistance. Probably the simplest example of such a situation is a Hagen-Poiseuille flow through a channel with an arbitrary slip length, where the average velocity is \emph{proportional} to the slip length. In particular, the velocity turns infinite in the no-stress limit where the slip length diverges.\\
This situation can be regularised if one takes into account weak momentum relaxation due to momentum-non-conserving impurities, phonons and Umklapp scattering. In order to do that one adds an Ohmic term proportional to velocity to the Navier-Stokes equation. In that case, however, the conductivity is dominated by the Ohmic rather than the viscous effects for large slip. This leads us to the question if one can somehow slow down the fluid near the boundary to make the viscous effects more pronounced.

\begin{figure}[hbtp]
\centering
\includegraphics[width=0.47\textwidth]{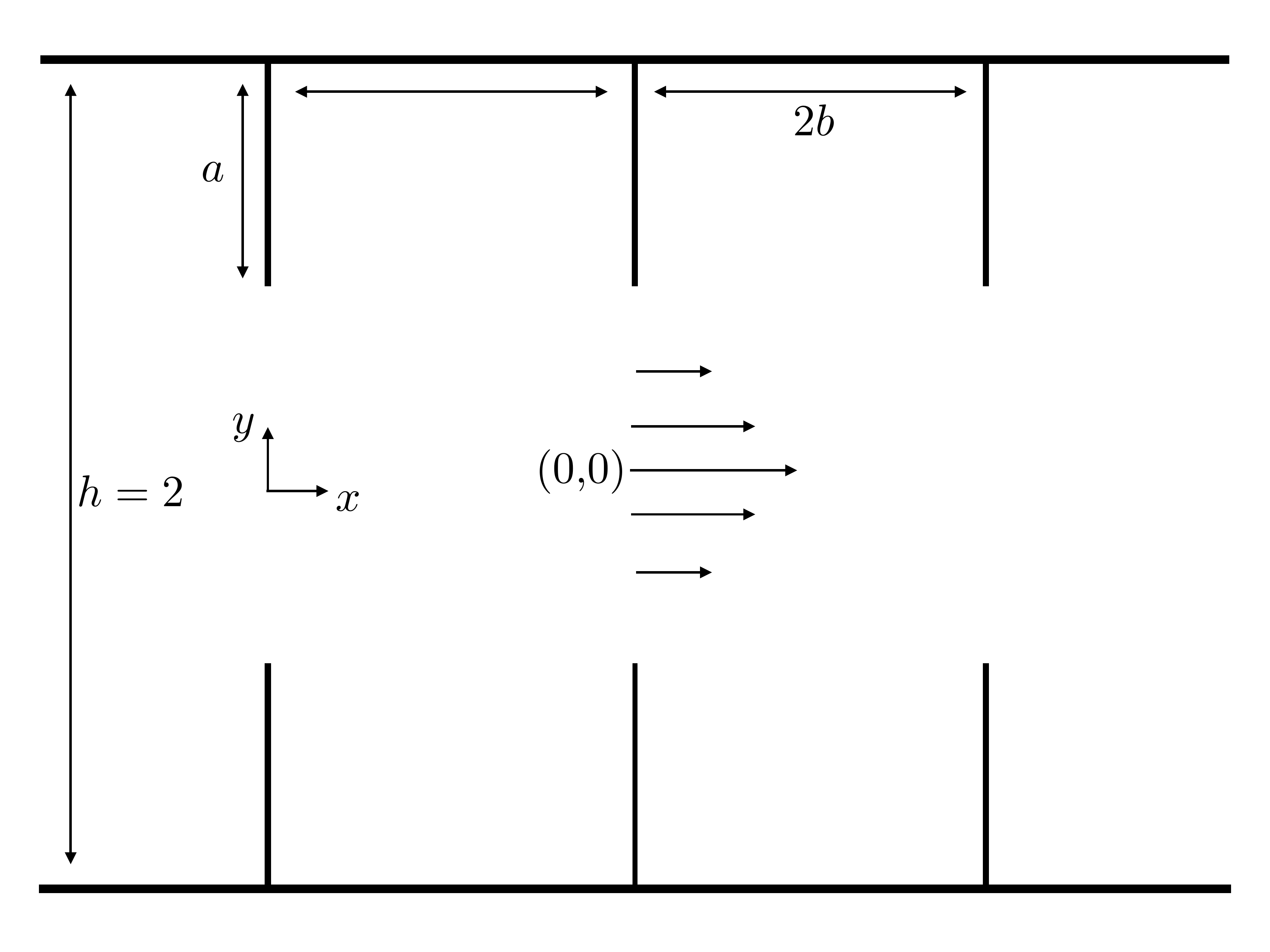} 
\caption{Flow geometry with a series of obstacles forming  cavities on the boundaries}
\label{pic:Channel}
\end{figure}

Our simple proposal is to slow down the fluid near the walls  by introducing a series of small obstacles on the boundary (Fig. \ref{pic:Channel}). The mechanism guaranteeing efficiency of that method takes up an idea by Moffatt, \cite{Moffatt1964} who noteced that an arbitrary viscous flow outside of a cavity will drive a vortical flow inside (see also \cite{Pan1967,Davis1993,Branicki2006} for a discussion of this effect in various set-ups). Later Wang \cite{Wang1997} constructed a solution for a Stokes flow with no-slip boundary conditions in a channel with perpendicular barriers equally spaced on the channel boundaries. He observed Moffatt vortices appearing inside the cavities below a critical distance between barriers.\\ 
Crucially, the induced vortices are typically tiny -- the flow velocity around such a vortex is orders of magnitude smaller that in the main driving flow. Thus, in general the fluid inside the cavity flows with a relatively small velocity compared to flow in the middle of the channel, thus mimicking the no-slip boundary.

To test this idea, we conduct a series of simulations of an infinite channel with a periodic array of obstacles on the boundary (see Fig. \ref{pic:Channel}). We calculate flow profiles with arbitrary slip parameters and deduce that for some range of obstacle lengths and spacings, there is no strong dependence of the flow profile on the slip parameter, and indeed the flow in the center of the channel resembles a standard no-slip parallel Poiseuille flow. To additionally check our results, we repeat the simulations for a periodically driven AC flow, and observe the development of a boundary layer in the high frequency regime. That phenomenon, like the Gurzhi effect, is characteristic of the viscous flow regime \cite{Moessner2018}, but it is absent in parallel flows with no-stress boundary condition.

In our computation, which is performed for two-dimensional systems, it is convenient to use a stream function formulation of the flow. The most general time-dependent case is governed by the following equation
\begin{align}
    \partial_t\,\Delta\Psi-\eta\Delta^2\,\Psi+\gamma\Delta\Psi=0.
\end{align}
Here $\eta$ is the viscosity and $\gamma$ the coefficient of the Ohmic term. We concentrate first on the time-independent (DC) case. The equation can be simplified
\begin{align}
\Delta^2\,\Psi-\Gamma\Delta\Psi=0,~~\Gamma=\frac{\gamma (h/2)^2}{\eta}.
\end{align}
In the equation above, the spatial variables are dimension-less.
 In order to study the behavior of the flow through the finned channel, we define parameters 
\begin{align}
    \sigma \equiv \frac{2a}{h} \hspace{1cm} \text{and} \hspace{1cm} \beta \equiv \frac{2b}{h},
\end{align}
$\sigma$ being a fraction of the channel in which the fluid is blocked, so in the middle we have a free 'channel' of width $h(1-\sigma)$, while $\beta$ measures the aspect ratio of the unit cell. 

\begin{figure}
    \centering
    \includegraphics[width=0.9 \linewidth]{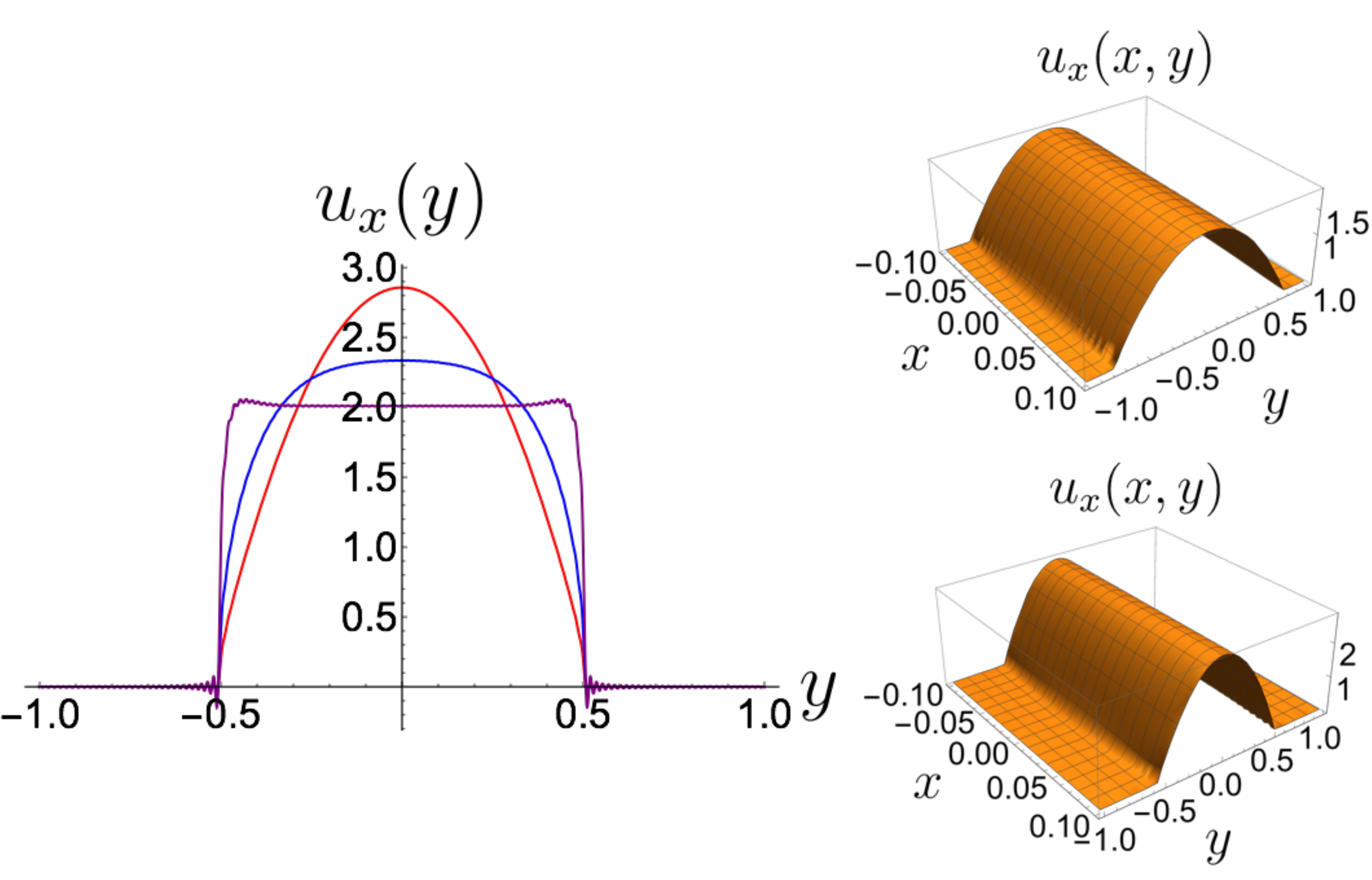}
    \caption{{\bf Left panel}: Flow velocity cuts at the edge of the unit cell, $x=1/10$, for different Ohmic coefficients. The three plots correspond to no-slip boundary conditions and parameters $\sigma=1/2$, $\beta=1/10$ (see text). Red line $\Gamma=1$ (hydrodynamic), blue line $\Gamma=100$ (crossover) and purple line $\Gamma=1000$ (ohmic). {\bf Right panel}: Profile of the flow velocity along a unit cell for no-stress boundary conditions, both plots correspond to $\beta=1/10$, while $\sigma=1/4$ (top) and $\sigma=1/2$ (bottom).}
    \label{pic:Static}
\end{figure}

In Fig. \ref{pic:Static} (left panel), the $x$-component of the velocity along the line $x=1/10$, for a unit cell with $\beta=1/10$ and $\sigma=1/2$, is plotted for different Ohmic coefficients. It can be seen that it looks parabolic for small Ohmic dissipation, resembling hydrodynamic behavior. For higher values of the Ohmic coefficient the flow profile becomes flat. This result is for no-slip boundary condition and is in correspondence with \cite{Guo2017}.

 More interestingly, if the no-stress boundary condition is implemented, the flow through the middle aperture of the channel still resembles a parabolic flow, as can be seen from the right panel of Fig. \ref{pic:Static}. This is our first central result.\\\\

\begin{figure}
    \centering
    \includegraphics[width=0.99 \linewidth]{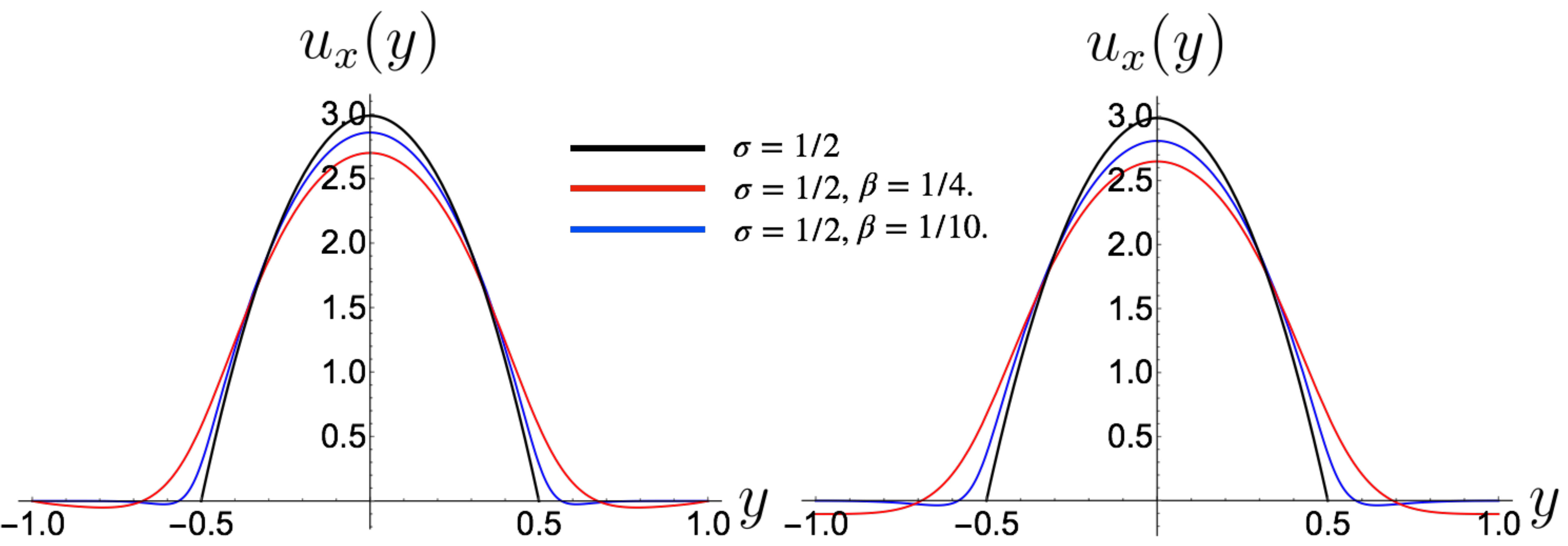}
    \caption{Flow velocity cuts at the center of the unit cell, $x=0$, for no-slip (left) and no-stress (right) boundary conditions. The black lines correspond to Poiseuille flows for an effective center channel of width $\tilde{h}=1/2=1-\sigma$. Recall that we are in the low dissipation regime, $\Gamma=1$. Note the similarity between the actual flow and the Poiseuille flow.}
    \label{pic:Static2}
\end{figure}

To investigate the similarity of the flow with the Poiseuille case in detail, we fix the dimensionless Ohmic coefficient $\Gamma=1$, which places it in an experimentally feasible range\cite{Bandurin2016,Moll2016,Mackenzie2017}. In Fig. \ref{pic:Static2} we plot how the velocity profile in the middle of the channel changes with respect to the parameter $\beta$ for both no-slip and no-stress boundary conditions.  All plots are for fixed $\sigma=1/2$, therefore we also plot the velocity for the Poiseuille flow on a flat channel of width equals one for comparison. 

\begin{figure}
    \centering
    \begin{tabular}{c | c | c}
    $\beta=1/2$ & $\beta=1/4$ & $\beta=1/6$\\
    \includegraphics[width=0.3\linewidth]{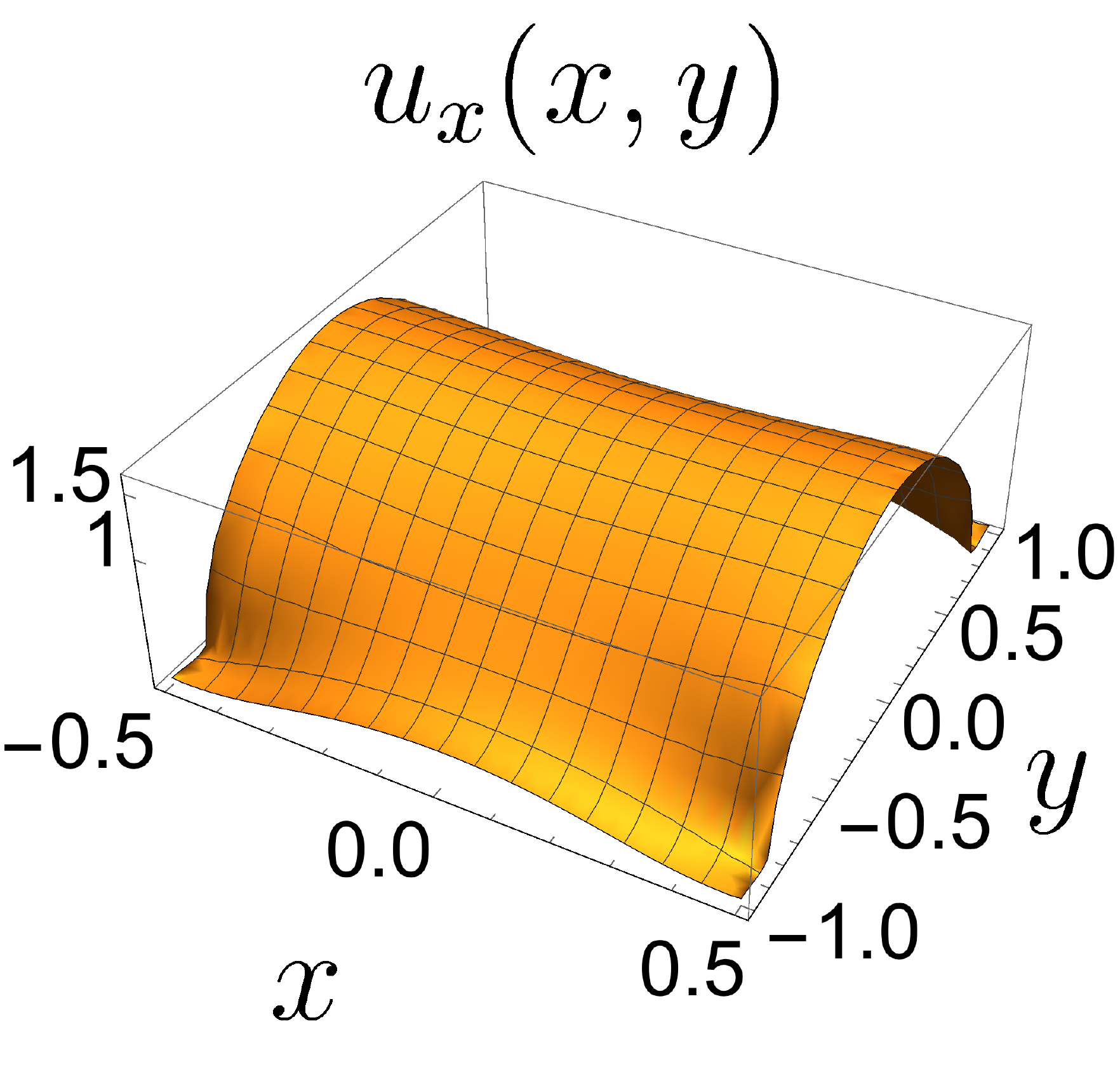} & \includegraphics[width=0.3\linewidth]{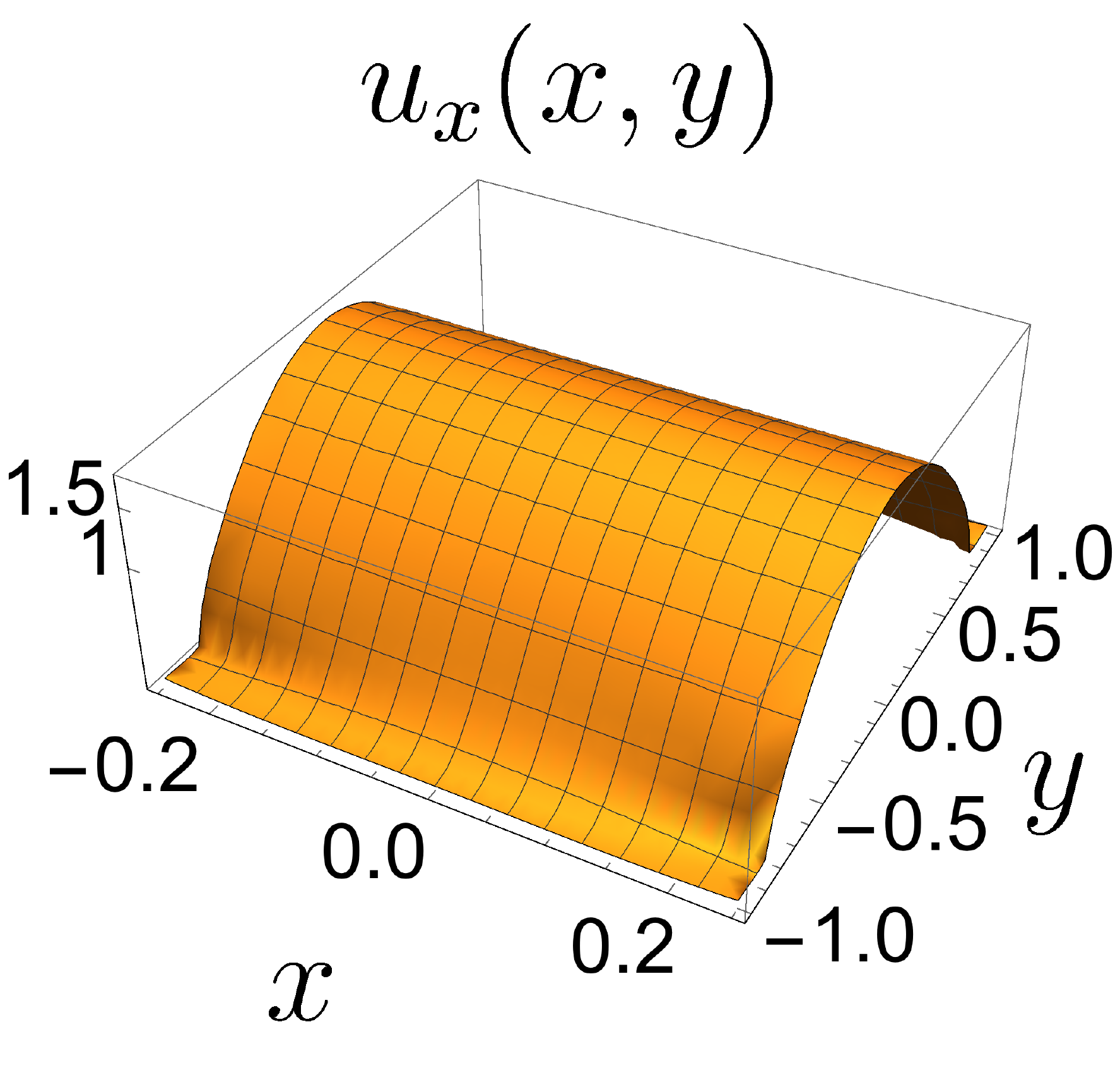} &\includegraphics[width=0.3\linewidth]{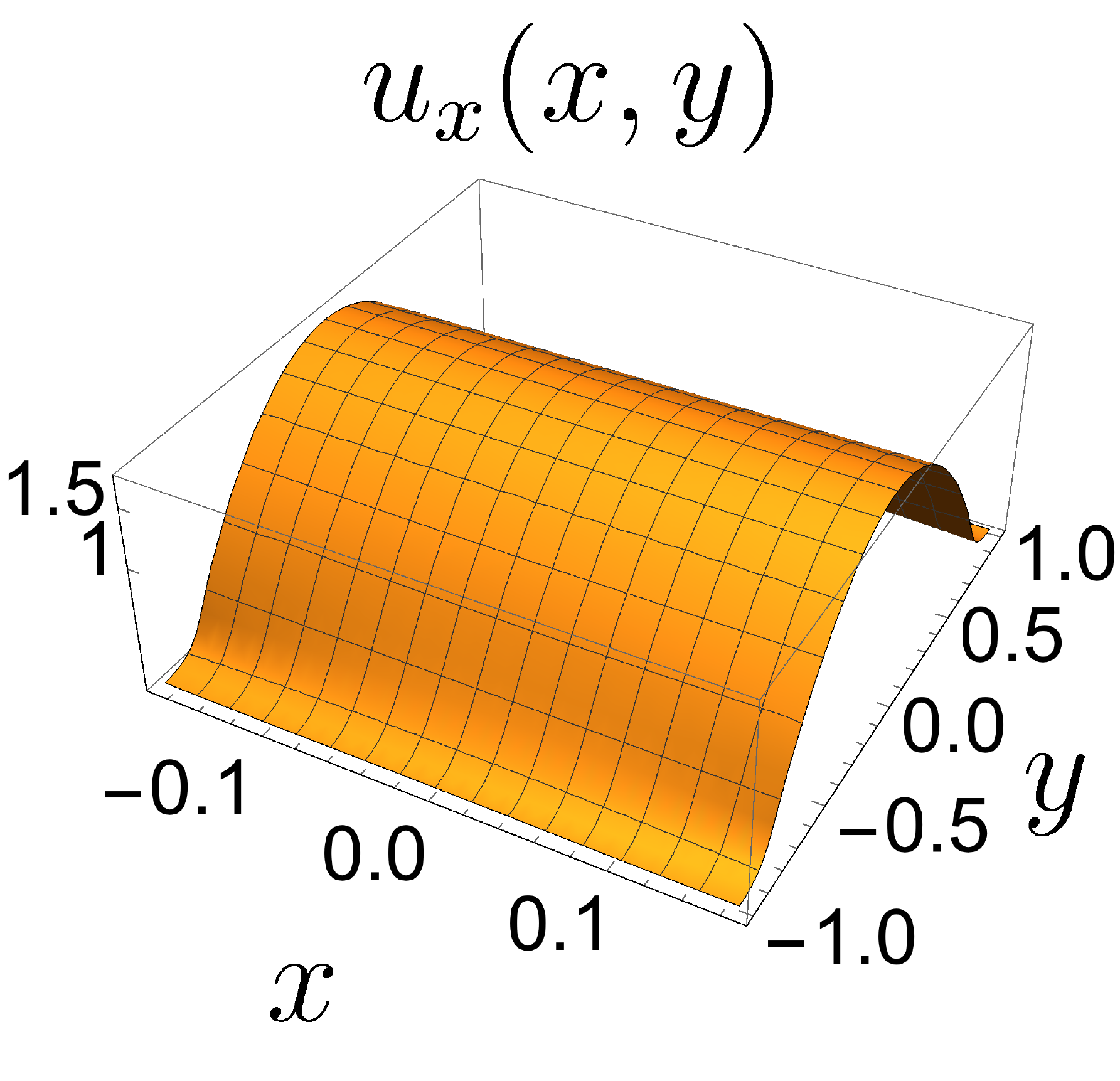} \\ 
    \includegraphics[width=0.3\linewidth]{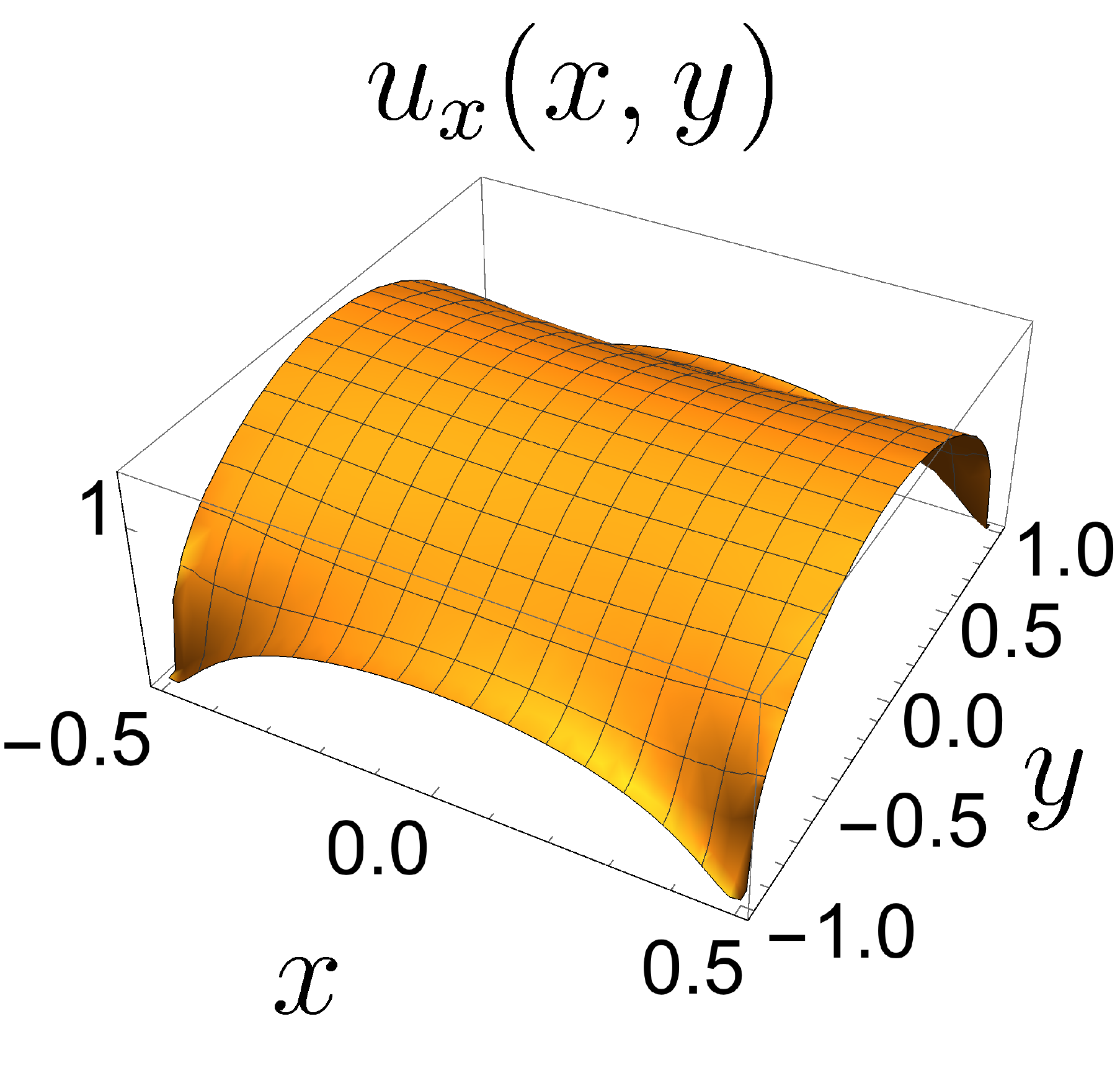} & \includegraphics[width=0.3\linewidth]{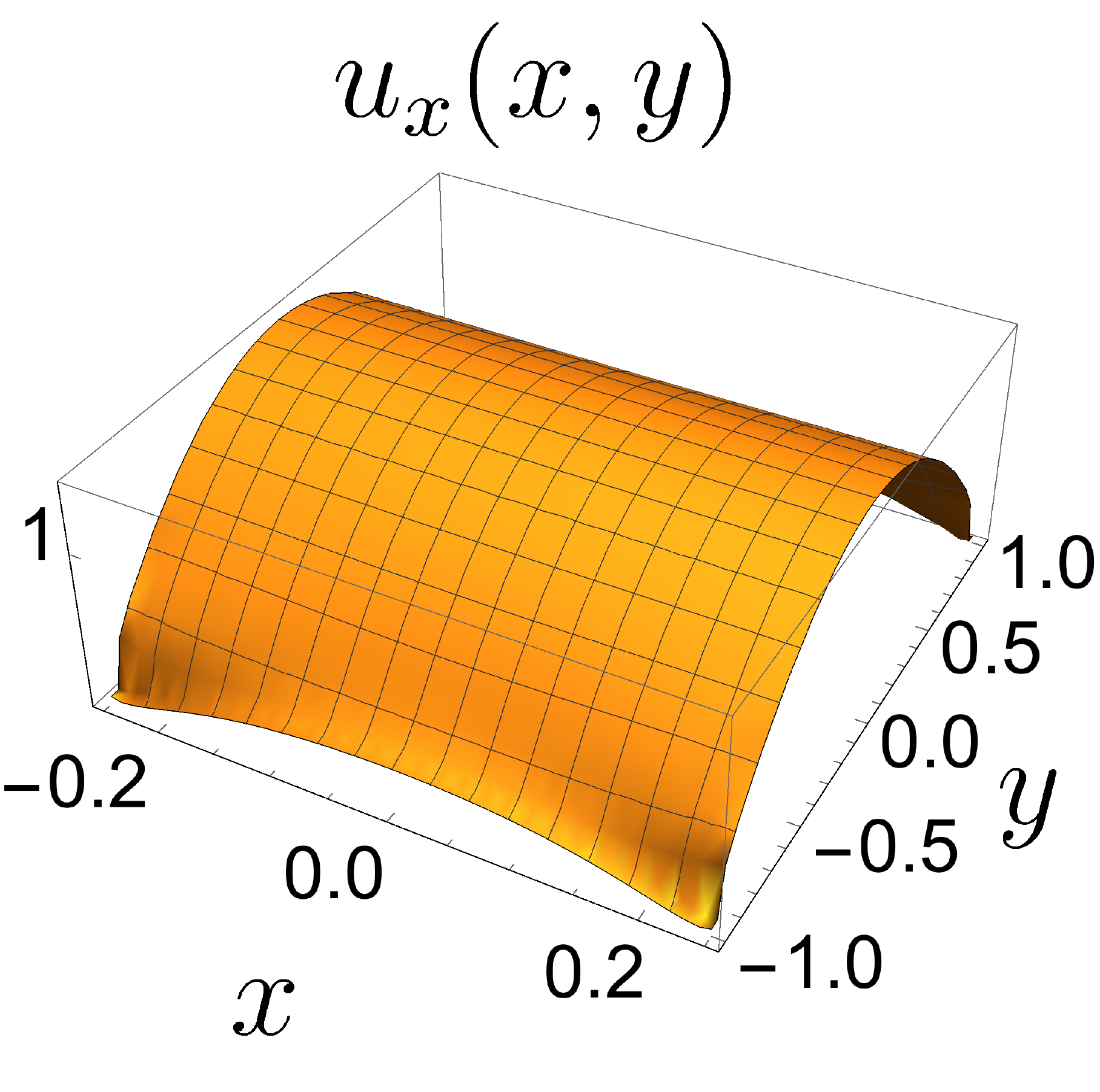} &\includegraphics[width=0.3\linewidth]{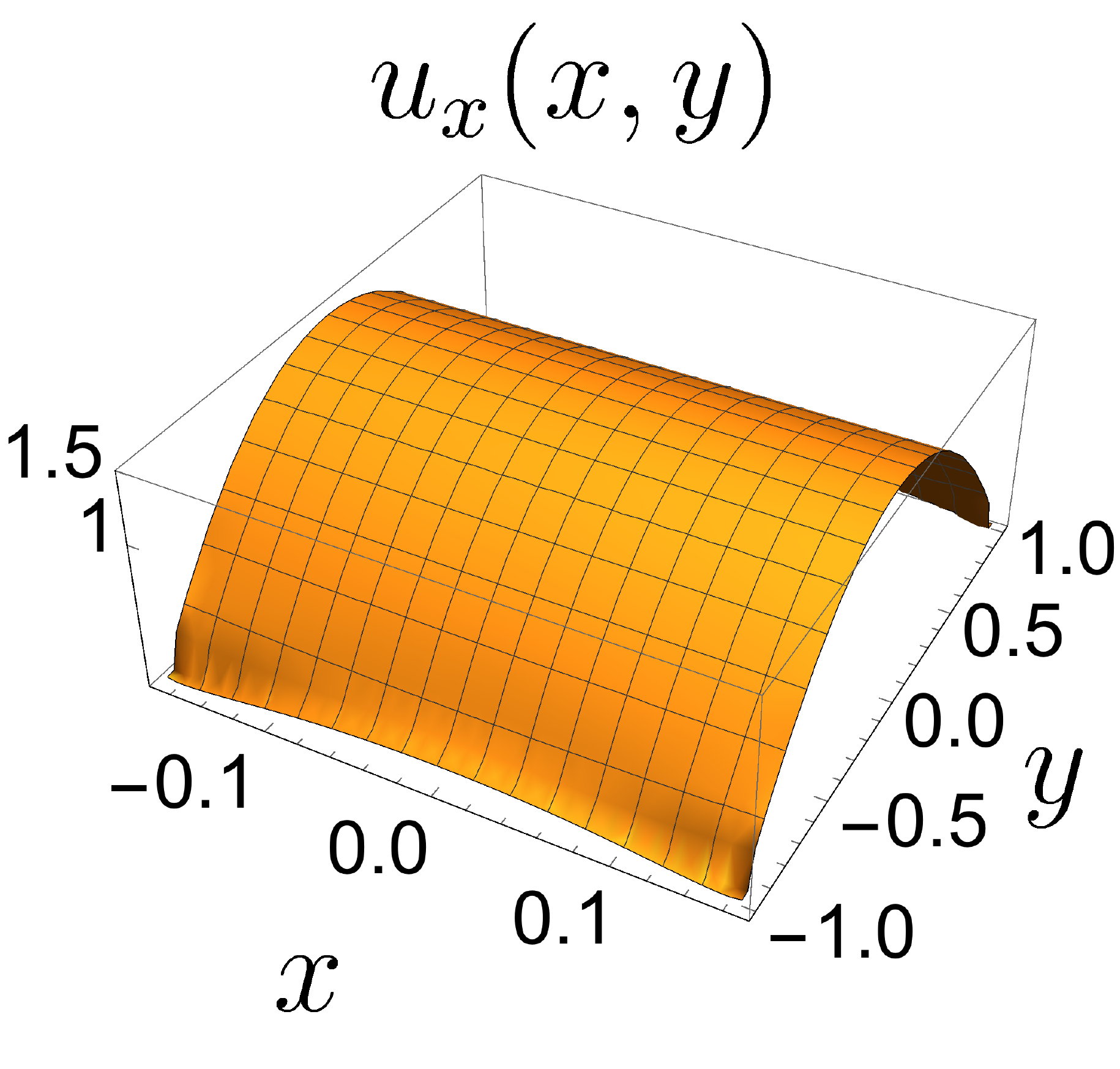}\\
        \includegraphics[width=0.3\linewidth]{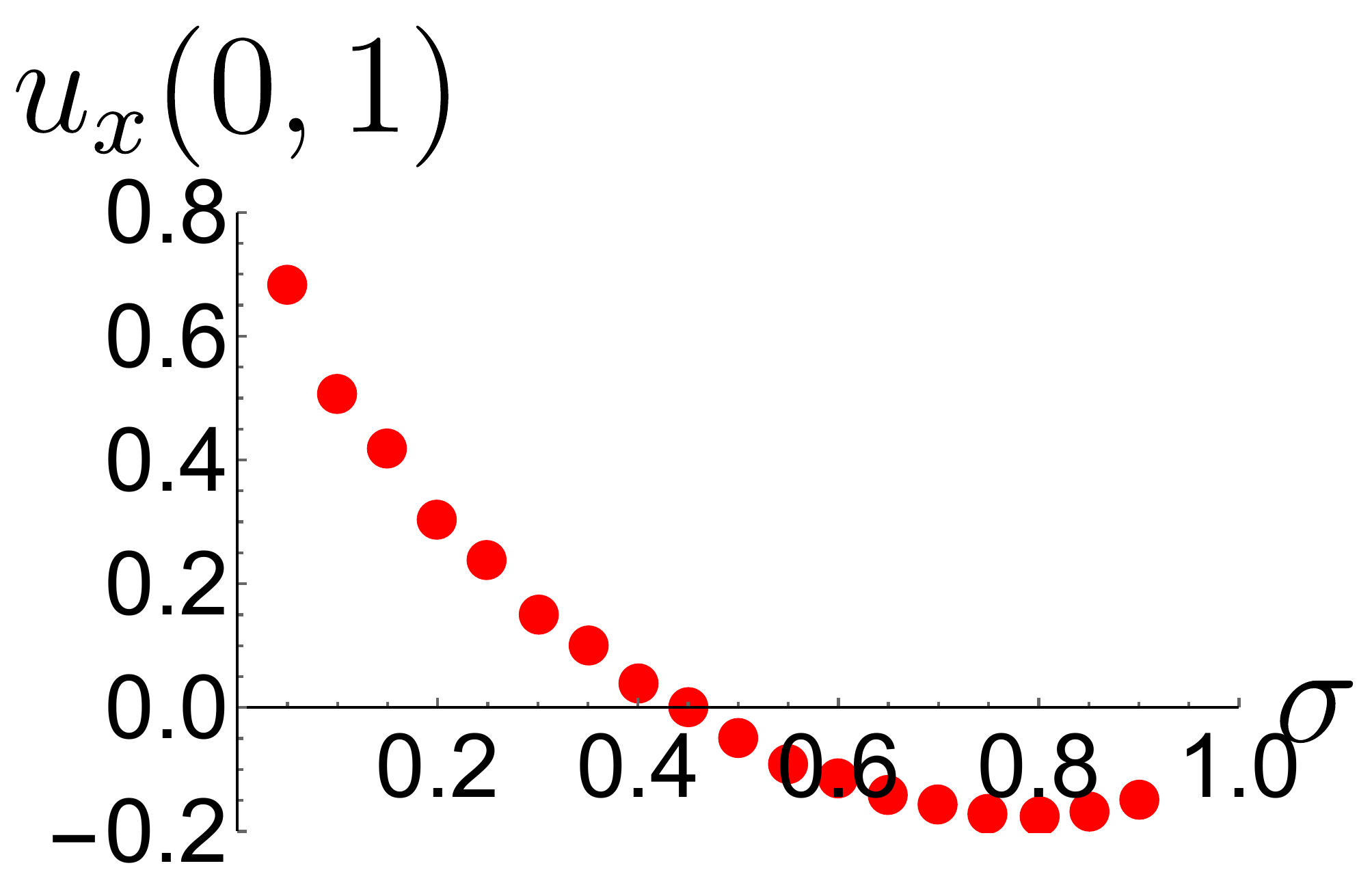} & \includegraphics[width=0.3\linewidth]{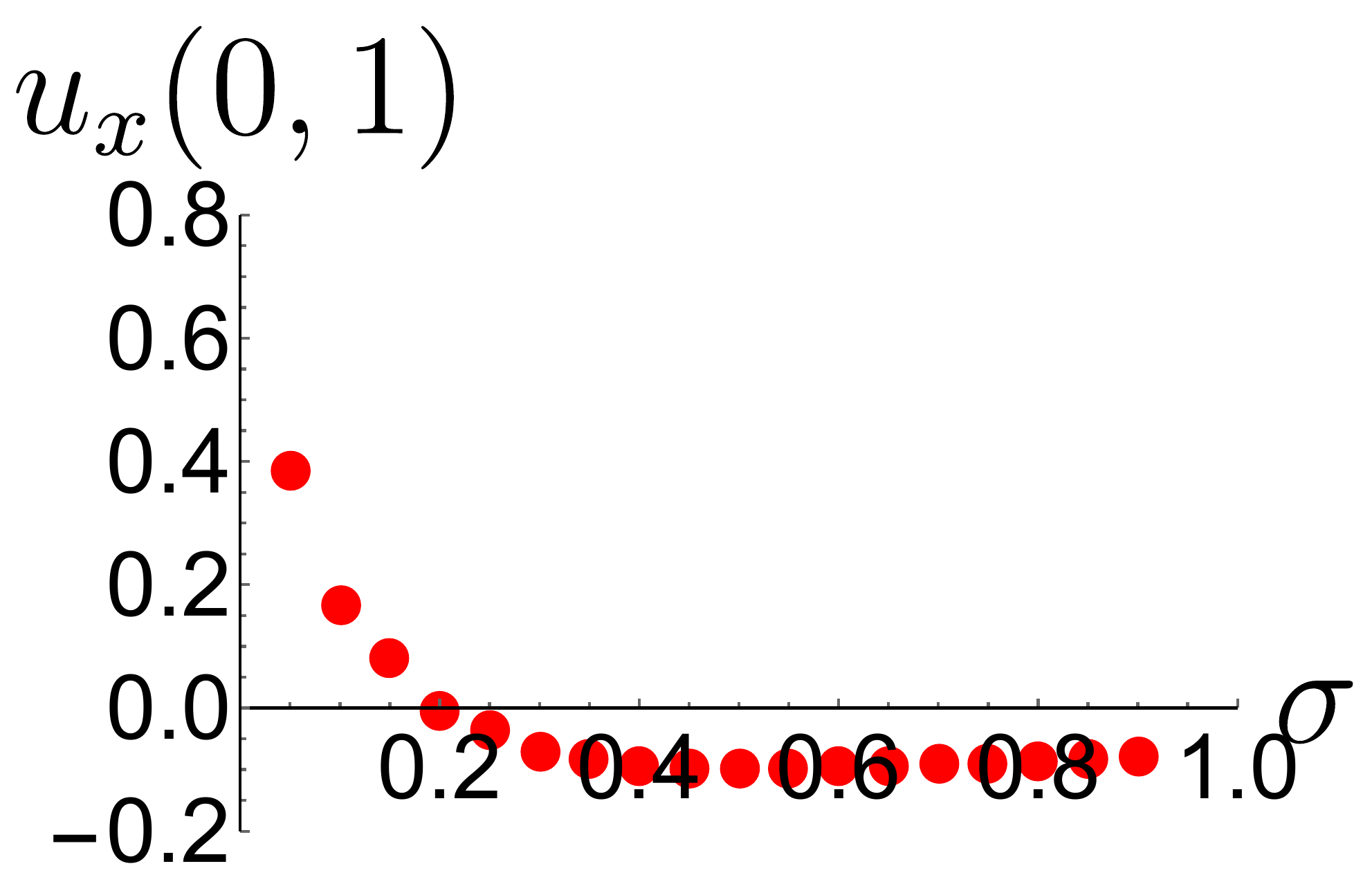} &\includegraphics[width=0.3\linewidth]{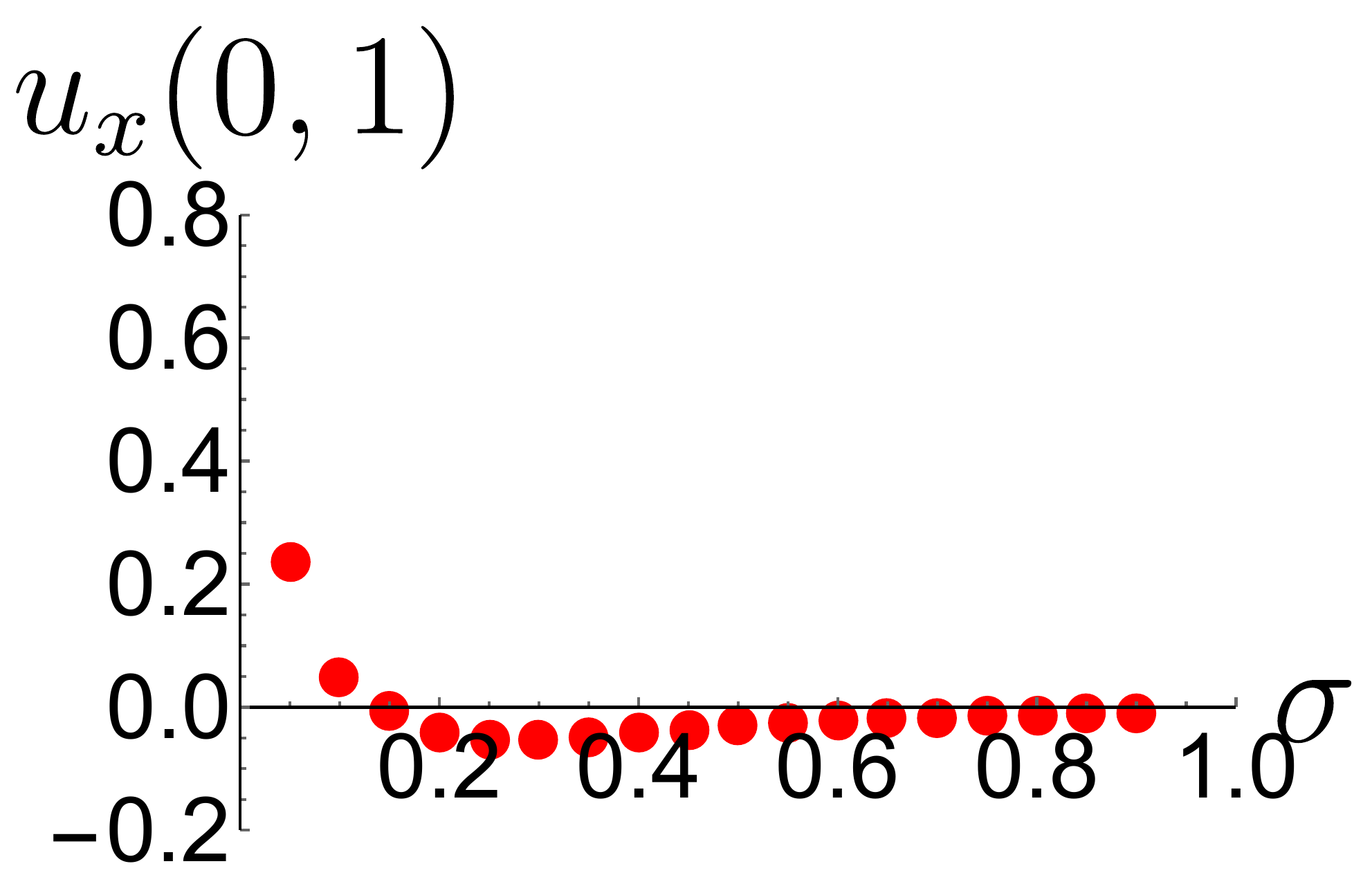} 
    \end{tabular}
    \caption{{\bf Top}: Velocity profiles along the channel for the corresponding values of $\beta$ and $\sigma=1/5$, corresponding to small obstacles in comparison with the height of the channel. {\bf Middle}: Velocity profiles along the channel for the corresponding values of $\beta$ and $\sigma=1/20$, corresponding to very small obstacles in comparison with the height of the channel, $h=2$ in our units. {\bf Bottom}: Value of the flow velocity at the top wall, coordinates $(x,y)=(0,1)$, of a unit channel cell as a function of $\sigma$ and for three values of $\beta$.}
    \label{pic:Static3}
\end{figure}

Given the similarity of the obtained profiles with the parabolic flow, we define an effective channel 
in the middle of our sample. Along this effective central channel, the fluid behaves as if the boundary condition on its walls were no-slip, regardless of the actual boundary conditions on the full finned channel.\\\\               
How small can the obstacles be made for the flow along the effective channel to be effectively parabolic? This is addressed in Fig. \ref{pic:Static3}, which depicts the evolution of the velocity profile for different values of $\beta$, and as a function of the aperture parameter $\sigma$. In the top and middle panels of Fig. \ref{pic:Static3}, the spatial dependence of the flow velocity is plotted for different parameters $\beta$ and $\sigma$, concentrating particularly on small obstacles. For some combinations of these parameters the flow closely resembles the parabolic no-slip flow.  Other combinations of parameters, corresponding to distantly spaced obstacles, yield flows differing considerably from classical Poiseuille.
%\begin{multicols}{2}
%\\\\
%
\def \combinesize {width=0.3\linewidth}
%\end{multicols}

In Fig. \ref{pic:Static3} (bottom panel), the velocity on the no-stress wall is plotted in red as a function of $\sigma$ for $\beta$ fixed. We see that when the velocity at the wall is not zero (or near to zero), the flow profile in the middle of the channel differs qualitatively from the parabolic Poiseuille flow. We use this to propose a criterion indicating when the effective flow in the middle channel resembles a parabolic flow. The analysis of numerical data suggests, that in order to ensure that the velocity in the cavity is negligibly small the following condition needs to hold
\begin{align*}
    \beta \alt \sigma,  
\end{align*}
i.e. it is important that the obstacles are not too far away from each other -- the distance between them should be of order of obstacle length or smaller.

Note for example that for $\beta=1/10$ (the parameter of the plots in Fig. \ref{pic:Static}), there is a large range of values of the $\sigma$ parameter for which the velocity at the walls is very close to zero despite the no-stress boundary condition. This supports the idea that a series of obstacles can effectively change the slip parameter in a channel.

Also, note that mimicking no-slip boundary conditions reproduces more than just the simplest effects (such as parabolic Poiseuille flow). To confirm that fact, we repeat the above simulation in a time-dependent (AC) scenario with a periodic forcing.

\begin{figure}
    \centering
    \includegraphics[width=0.9 \linewidth]{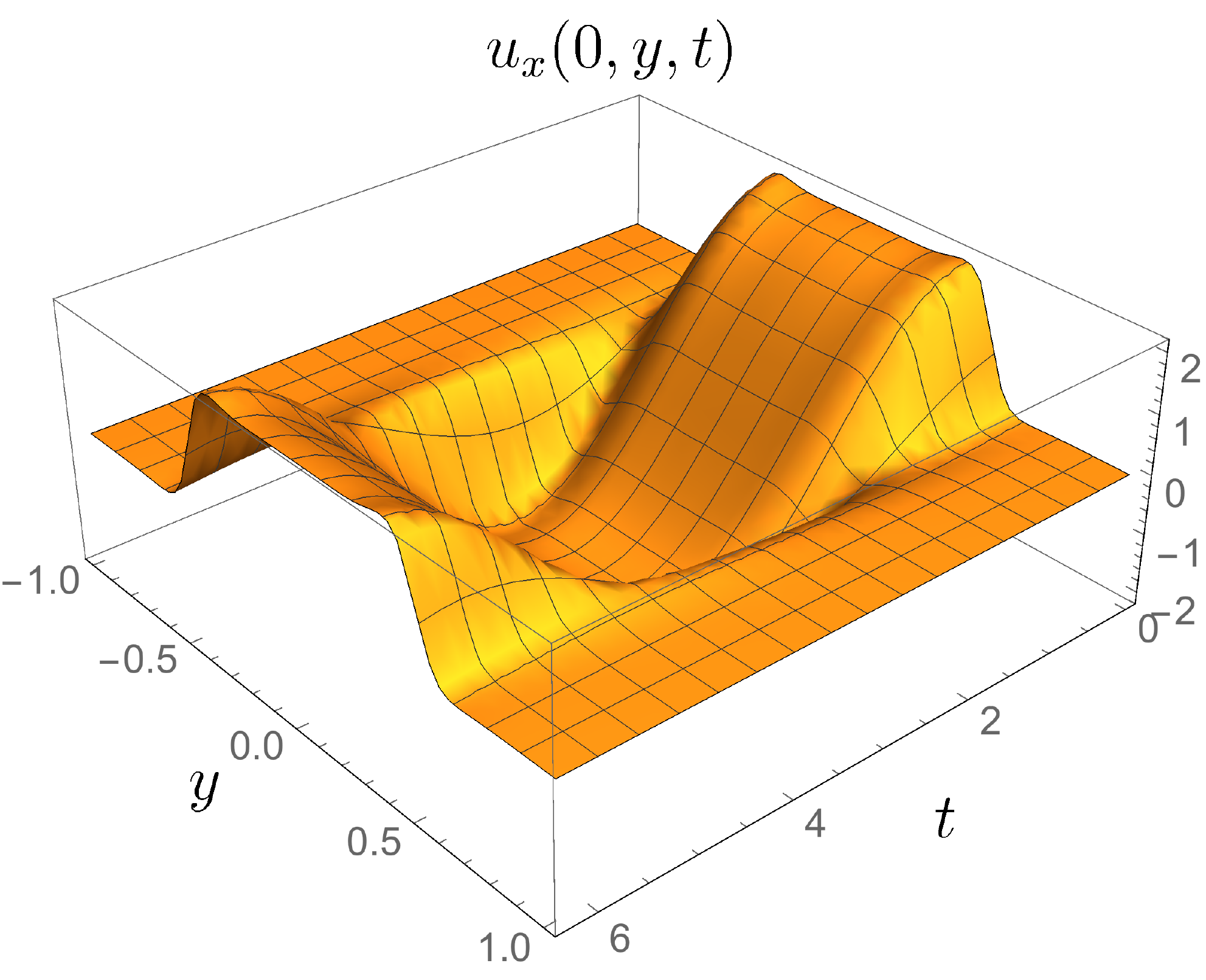}
    \caption{Spatiotemporal velocity cut profiles at $x=0$, for high frequency driving over one period, $t \in [0,2\pi]$. No-stress boundary conditions on walls and obstacles. Parameters: $\Lambda=500\, i+1$, $h=2$, $\sigma=1/2$ and $\beta=1/10$.}
    \label{pic:Time1}
\end{figure}
The Stokes boundary layers emerge above a certain forcing frequency (see Fig. \ref{pic:Time1}). This phenomenon, previously described for a flat channel\cite{Moessner2018}, is also tied to the no-slip boundary condition, in the following way.
For large frequencies, the fluid "cannot follow" the drive, and stops to be in phase with the rapidly oscillating force. The fluid in the middle of the channel oscillates uniformly, and only close to the boundary does the viscosity become important. This has to do with the frequency of forcing being too big for the viscosity to efficiently transport the momentum through the whole channel. As a consequence, a strong gradient is created near the boundary, on a distance that corresponds to the effective 'range' of viscous interaction under periodic driving. This gradient of course only emerges if the fluid sticks to the boundary, i.e the velocity there is close to zero. 

Since in parallel to the flow in channels without barriers, fast forcing in our set-up results in the maximal velocity at some distance from the center, we conclude that the structured boundary indeed does mimic a no-slip boundary quite well. Apart from that, this analysis supplements existing literature on time dependent electronic flows \cite{Tomadin2014,Semenyakin2018,Alekseev2018}.
\section{Flows with curved boundary}\label{sec:curvedbdr}
\subsection{Boundary conditions on curved geometry} 

The behavior of a fluid flow at the interface with other bodies (i.e. on the boundary of the sample) is a complicated one that crucially influences the solutions of the theory. Plenty of non-trivial physical phenomena governing this behavior are contained in effective descriptions in terms of a proper boundary condition \cite{Richardson1973,Hocking1976,PanzerPRL,panzer1992effects,Tuck1995,Sarkar1996}. Indeed, various characteristics of our system modify the slip length. They include temperature and parameters related to the wall material and fluid composition, as well as mesoscopic and sub-mesoscopic components, in particular the wall curvature \cite{Panton}. A quantitative understanding of the wall curvature in terms of an effective slip value was given by Einzel, Panzer and Liu\cite{PanzerPRL}:
\begin{equation}\label{eq:slipeff}
\xi_\text{eff}= \left( \frac{1}{\xi_0}-\frac{1}{R}\right) ^{-1},
\end{equation}
where $R$ is the curvature radius measured in such a way, that it is positive if the fluid domain is convex and negative otherwise, see Fig. \ref{pic:CurvConventions}. For an explanation of this condition see Appendix \ref{sec:EPL}.
The fact that the boundary curvature modifies the slip parameter has a direct influence on solutions on more complicated domains.  On top of that one needs to consider sub-mesoscopic roughness of the boundary, too big to directly influence boundary scattering of individual carriers, but not small enough to be approximated with a straight line on the scale of a system. Instead the roughness modifies the slip length with an effective curvature contribution\cite{panzer1992effects}.
In this section we focus on the effective description of curvature effects in the context of electronic fluids. 
\begin{figure}[hbtp]
\centering
\includegraphics[width=.9\linewidth]{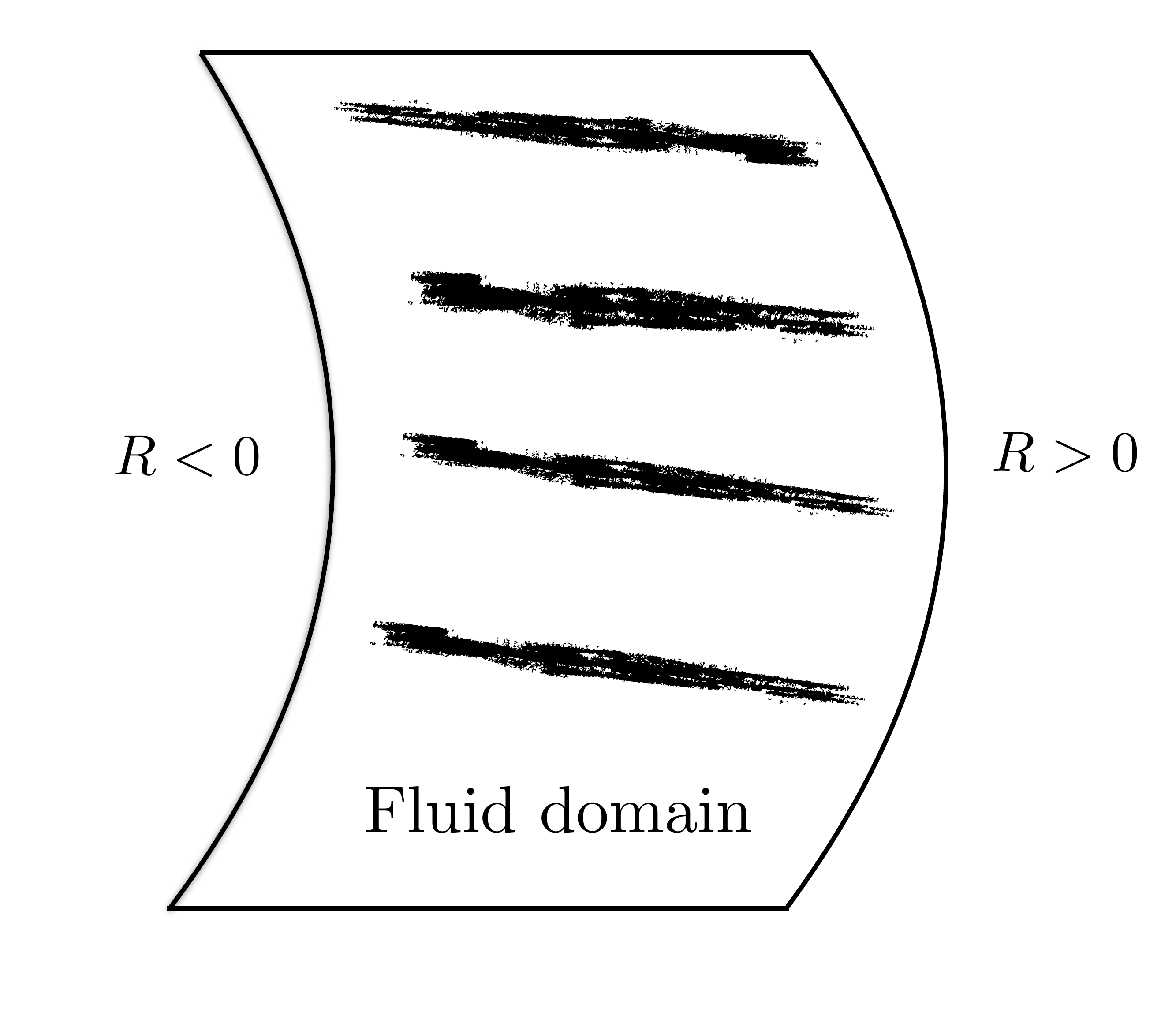}
\caption{The conventions on boundary curvatue used in \ref{eq:slipeff}. $R>0$ for convex domain boundary, $R<0$ for concave.}\label{pic:CurvConventions}
\end{figure}

  Having introduced how the boundary condition gets  modified, we proceed to test experimental implications in a couple of set-ups.

\subsection{Flow in a channel with mesoscopic boundary curvature} The analysis above shows that geometric effects can have crucial impact on the slip-length. In materials like graphene, the sample production process may yield a boundary that is rough not only on the microscopic scale but at scales up to the sample size, in which case the boundary curvature needs to be taken into account. 
As a simple, yet very instructive example, we take a channel that is almost flat. More precisely,  we study the curvature radii on both sides of the sample $R_1$ and $R_2$ which are much bigger than the channel width $w$. If the microscopic slip length is also large compared to that scale, we can approximately use the flat channel parallel flow solution in which the curvature effects manifest themselves in the modified slip length $\xi _{\text{eff}}$. We then have to solve the Stokes equation
\begin{equation}\label{eq:Stokes}
\eta \frac{\partial ^2 u_x}{\partial y^2}=-\frac{e}{m}\frac{\partial \phi}{\partial x},
\end{equation}
where $\eta$ is the viscosity coefficient and $\phi$ is the electrochemical potential. The most general situation allows the curvature radii to be different on the two boundaries of the channel:
\begin{align}\label{eq:bc2}
	  \left.u^t_i\right|_{B_1}=\left.\xi_1\,n_j\,\frac{\partial u^t_i}{\partial x_j} \right|_{B_1}, \\ 
	%%%%%%%%%%%%%%%%%%%%%%%%%%%%%%%%%%%%%%%%%%%%%%%%%%%%%%%
	 \left.u^t_i\right|_{B_2}=\left.\xi_2\,n_j\,\frac{\partial u^t_i}{\partial x_j} \right|_{B_2},
\end{align}
where the boundaries are located at $\{-w/2,w/2\}$. The above boundary conditions lead to the following velocity profile
\begin{equation}\label{eq:velocity}
u_x(y)=\frac{1}{8\eta (w+\xi_1+\xi_2)} \mathcal{U}(y) \frac{e}{m}\frac{\partial \phi}{\partial x},
\end{equation}
where
\begin{align}
\mathcal{U}(y) &=w^3 -4 w y (y+\xi_1)+4w(y+2 \xi_1)\xi_2 \\
                        &+3 w^2 (\xi_1+\xi_2) -4y^2 (\xi_1+\xi_2)  .\nonumber
 \end{align}
Integrating this expression we obtain the total current
\begin{align}\label{eq:current}
I&=\int  _{-w/2}^{w/2}dyu_x(y)\\ \nonumber
&=\frac{w^2[w^2+12\xi_1 \xi_2 +4w(\xi_1+\xi_2)]}{8\eta (w+\xi_1+\xi_2)} \frac{e}{m}\frac{\partial \phi}{\partial x}.
\end{align}
We see that the Gurzhi effect corresponds to an idealised situation, with zero slip, legitimate only in set-ups where both the microscopic contribution to the slip-length as well as the geometric component are much smaller than the width of the channel.

Since we are interested in the role of slip on the electronic flow, we use this solution to investigate geometric contributions to the current. The importance of this analysis stems from the fact that in most experimental samples the flat channel serves as a theoretical benchmark, and the boundary conditions are an important missing ingredient for a more comprehensive theoretical understanding. 

\begin{figure}[hbtp]
\centering
\includegraphics[width=0.47\textwidth]{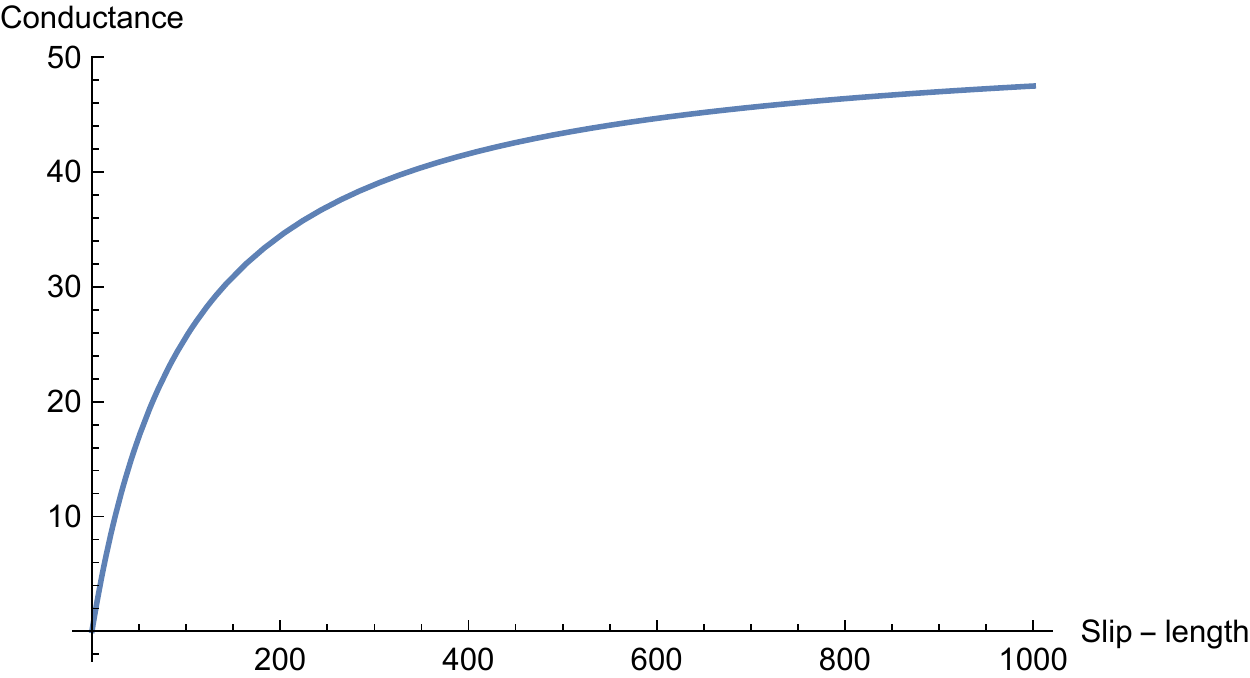} 
\caption{Conductance as a function of the slip-length for boundaries with positive curvature with $\eta=1$, $R_1=110$, $R_2=100$}
\label{pic:CondPosPos}
\end{figure}

In our set-up we distinguish three situations: the curvature is positive on both boundaries, the curvature is negative on both boundaries, and one boundary has positive and the other has negative curvature. We start with two boundaries with negative curvature (i.e. the channel is thinner in the middle). In this case
\begin{equation}\label{eq:pospos}
\xi_1= \left( \frac{1}{\xi_0}+\frac{1}{R_1}\right) ^{-1} , \qquad \xi_2=\left(\frac{1}{\xi_0}+\frac{1}{R_2}\right) ^{-1},
\end{equation}
where $\xi _0$ is the microscopic slip-length. Note, that obtaining the formal no-stress limit requires not only microscopic slip $\xi_0$ but also curvature radius $R$ to be infinite for positive curvature. We plot the corresponding curvature on Fig. \ref{pic:CondPosPos}.

The next example we consider is when the boundaries have both negative curvature
\begin{figure}[hbtp]
\centering
\includegraphics[width=0.47\textwidth]{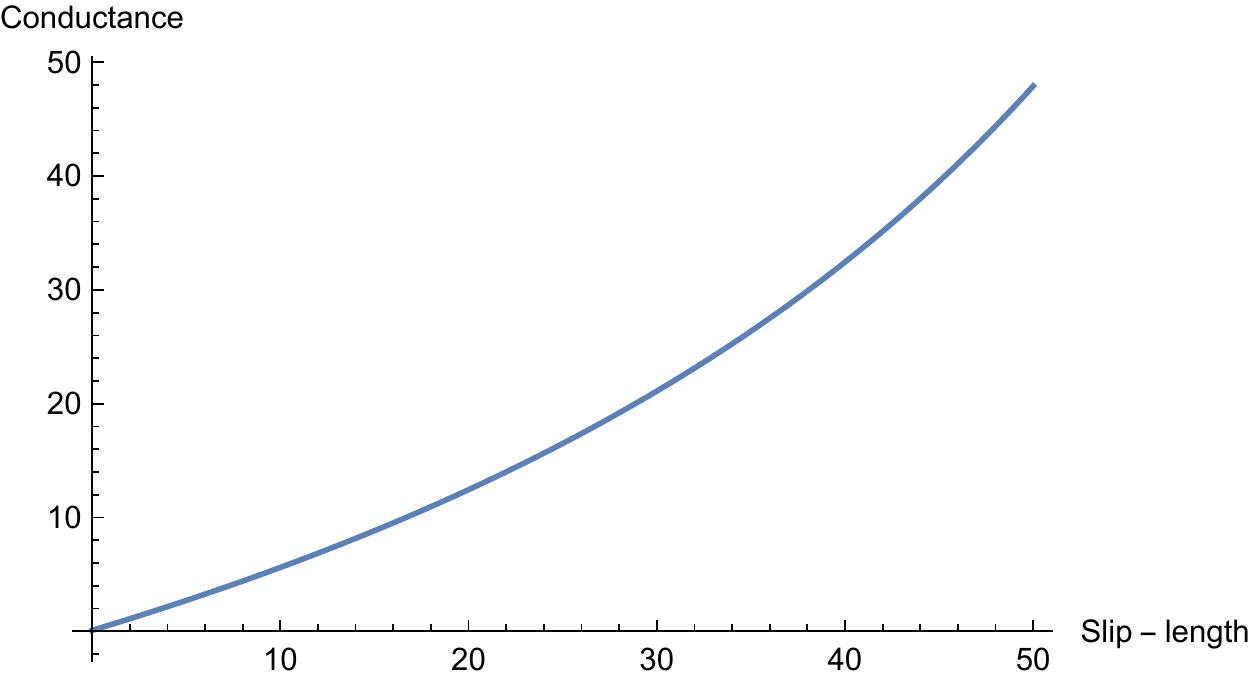} 
\caption{Conductance as a function of the slip-length for boundaries with negative curvature with $\eta=1$, $R_1=110$, $R_2=100$}
\label{pic:CondNegNeg}
\end{figure}

\begin{equation}\label{eq:negneg}
\xi_1= \left( \frac{1}{\xi_0}-\frac{1}{R_1}\right) ^{-1} , \qquad \xi_2=\left(\frac{1}{\xi_0}-\frac{1}{R_2}\right) ^{-1}.
\end{equation}
The conductance that follows changes character as can be seen on Fig. \ref{pic:CondNegNeg}. The negative contribution from geometry results in a special point where the microscopic and geometric slip-lengths are equal and an infinite jump of $\xi_1$ and $\xi_2$ happens. 
This is, however, not physical as in effect it leads to conductance having infinite jump with a sign change, yielding the negative conductance for some parameters.
Therefore, the boundary condition \eqref{eq:GeneralBCflat} is certainly unphysical when the curvature radius becomes comparable with the microscopic slip length. 

This problem also arises if only one side has negative curvature.

The main message of this section is that the boundary curvature changes many flow characteristics in fluids, which naturally have non-negligible slip-length. The resulting solution possesses a richer structure, which does not fall into two categories of either flat or parabolic flow profiles. The conductance that follows depends on two slip parameters, which contain the geometric characteristic of the boundaries. As a result it has neither a linear scaling with the channel width as in the ballistic regime, nor the quadratic scaling in the Hagen-Poiseuille regime. More precisely one can see that by looking at the solution \eqref{eq:velocity}, which has a form of the ratio of two polynomials and depends on two parameters. The choice of parameters can affect both shape and symmetry of the flow profile. 

As we mentioned before, in the cases with a negative slip-length, the solution turns unphysical yielding infinite jump of the flow velocity. Hence we need to ask a question: does the hydrodynamic theory itself provide some mechanism of resolving this infinity? One of the possible mechanisms of such a regularisation would be if the Poiseuille-type solution itself becomes unstable. Hydrodynamic instability means that the solution, although mathematically correct, is fragile and can be easily destroyed. On a more formal level, it means that there are some perturbations that, once introduced in the system, grow in time ultimately completely altering the nature of solution.\\
To check that possibility, we perform a linear stability analysis of the solution \ref{eq:velocity}. The analysis is based on the Orr-Sommerfeld equation for linear perturbations of a parallel flow \cite{orszag_1971,Spectral1,SchmidHenningson}. This analysis is technically involved so we describe it in detail in Appendix \ref{sec:stability}. Let us just remark, that such an analysis yields values of parameters (like Reynolds number Re) for which some linearised perturbation(s) around the base solution grow in time rather than decay or oscillate, and therefore fall under the definition of unstable perturbations.

It turns out that while positive slip stabilises the solution  \cite{He2008}, a negative value of the slip parameter renders a very unstable flow. In a channel with $w=2$ and the boundary slip lengths $\xi_1=\xi_2=-0.55$,  the flow is unstable even for extremely low Reynolds numbers Re=5. We conclude that the parallel flow approximation we employ becomes unreliable when thes microscopic slip length is large. This happens as a large microscopic slip length in combination with even slightly curved boundary (or a sub-mesoscopic roughness of the sample edge) can yield a negative effective slip that destabilises solutions drastically.
The importance of this result stems form the fact that the parallel channel  is often used as a benchmark geometry for hydrodynamic effects. Our result shows, that in viscous electronics where slip lengths may be large, results obtained in such a set-up must be approached with caution. 

\section{Flow through a circular junction} \label{sec:junctions}
\subsection{Flow profiles} Boundaries can modify the effective slip. Beyond the weak curvature considered above  we next study systems in which the curvature radius is smaller than the system size. In general, to extract the geometric contribution to slip, the solutions corresponding to any but the simplest set-ups become complicated and the resulting effect is not transparent. Therefore we investigate the flow through a circular junction. In such a set-up, the only geometric length scale is the circle radius, which is at the same time the radius of curvature of the boundary to be used in the effective slip \ref{eq:slipeff}. Technically, the high degree of symmetry allows one to separate variables and Fourier decompose the angular dependence. 

Finally, this set-up was also investigated in the ballistic regime, both experimentally and theoretically \cite{Marcus1992,Ishio1995,Schwieters1996}. 
A striking feature that emerges in the ballistic regime is that the conductance exhibits characteristic irregular fluctuations as a function of Fermi momentum. 

\begin{figure}[hbtp]
\centering
\includegraphics[width=0.47\textwidth]{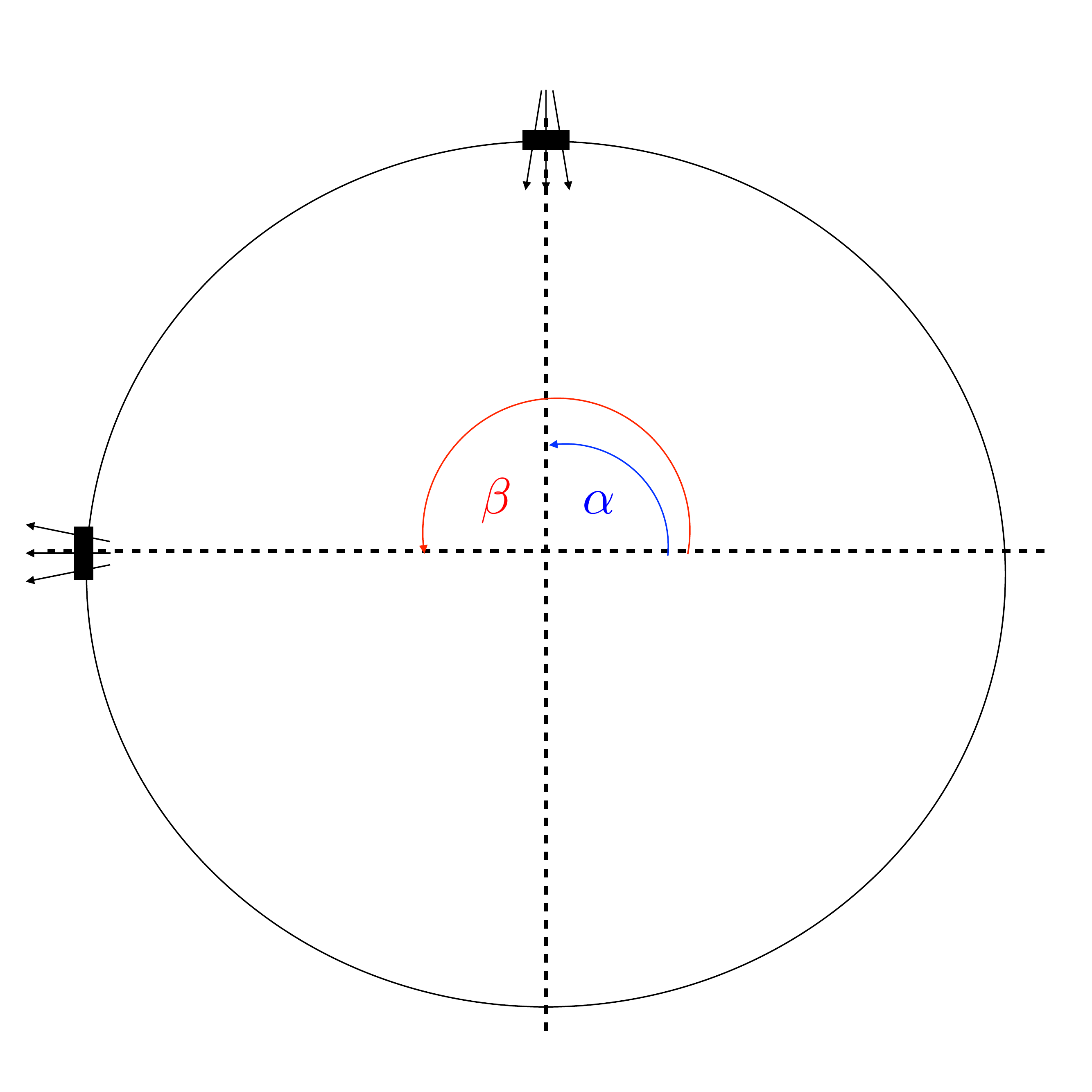} 
\caption{Inflow/outflow problem into a circular contact}
\label{pic:Disc}
\end{figure}

Our setup is a disc-shaped sample with two narrow contacts of width $\epsilon$.  The radial coordinate has range $r\in{}(0,1]$. The setup is presented in Fig. \ref{pic:Disc}

The problem of a fluid flowing into a circular domain through a boundary and then flowing out at some other boundary point has a long history in fluid literature. The solution can be constructed as a series expansion of the stream function \cite{Rayleigh1893,Mabey1957,Dennis1974,Mills1977,Dennis1993}. We relegate the details to Appendix \ref{sec:junctions}.
 \begin{figure}[hbtp]
 \centering
\begin{tabular}{c}\\
\includegraphics[width=0.40\textwidth]{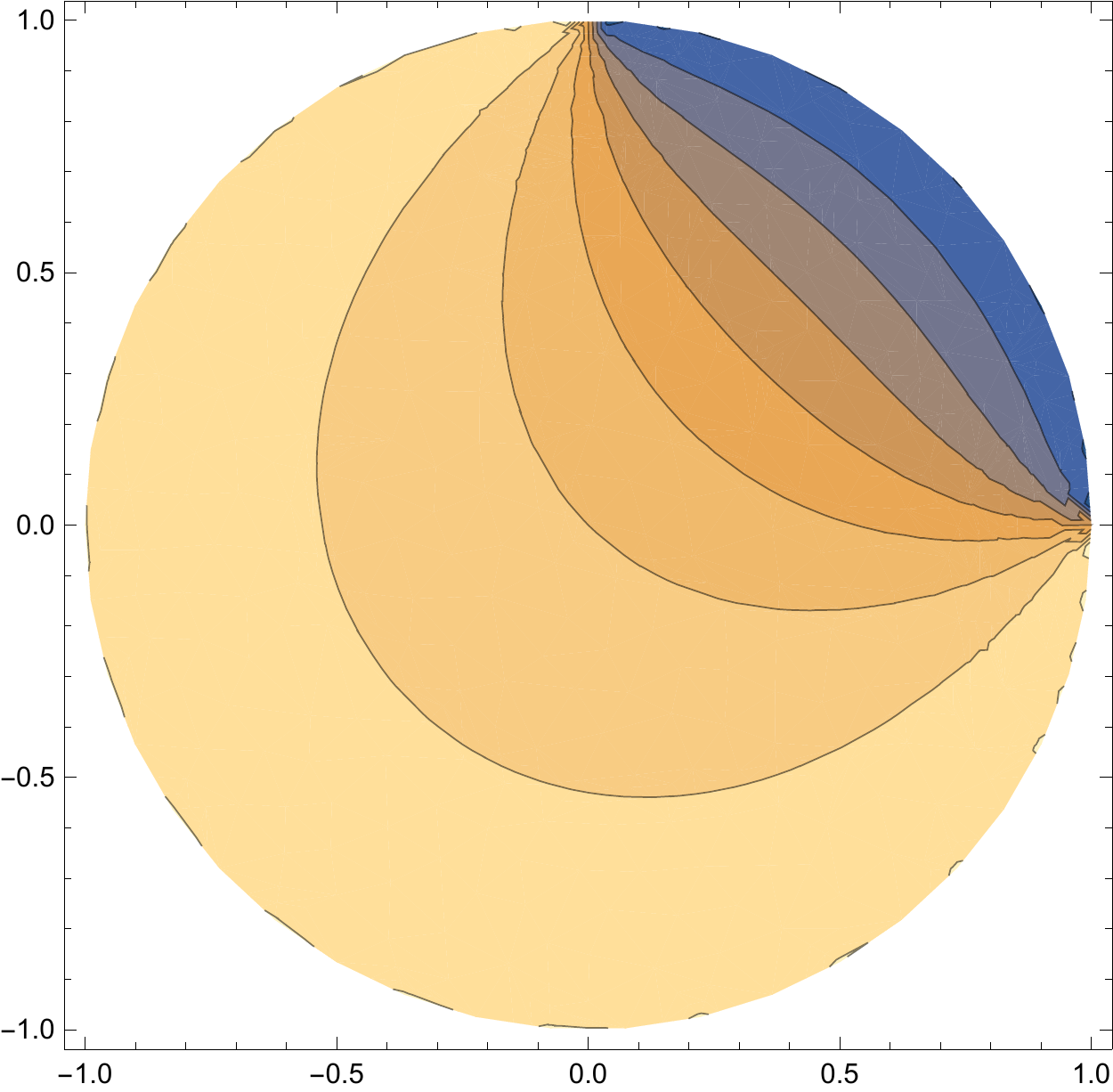} \\
\\
\includegraphics[width=0.40\textwidth]{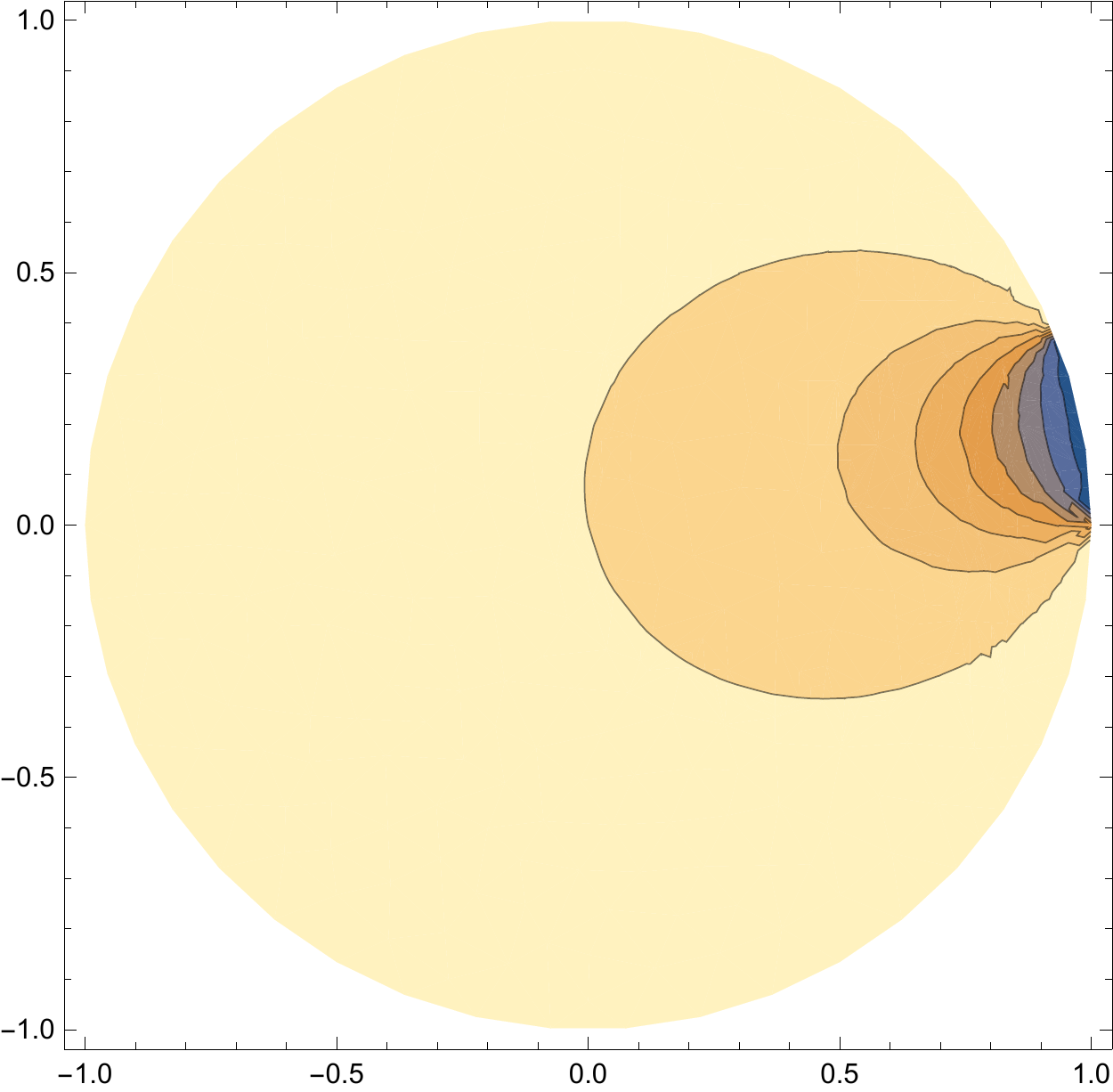} 
\end{tabular}
\caption{\textbf{Top:} Streamlines in a circular cavity between two contacts separated by an angle $\pi/2$ ($\alpha=0$, $\beta=\pi/2$). See \ref{pic:Disc} for the conventions on angles $\alpha,~\beta.$ \textbf{Bottom:} Streamlines between two contacts separated by an angle $\pi/8$ ($\alpha=0$, $\beta=\pi/8$).}\label{pic:CircularJunctions}
 \end{figure}

Junctions possess a big advantage over the channels, namely that a relatively simple theoretical analysis may be possible in both ballistic and hydrodynamic regimes unveiling the distinctive features. 
The hydrodynamic flow through a  confined geometry is smooth due to electron-electron interactions.  To illustrate this fact (see Fig. \ref{pic:CircularJunctions}) we plot a stream pattern for two configurations of contacts. In one configuration the contacts are separated by an angle $\pi/2$ in the second by $\pi/8$. Note that the former configuration was studied in the ballistic regime, both semi-classically\cite{Schwieters1996} and quantum-mechanically\cite{Ishio1995}. One can see that the flow profile is smooth. Moreover, the closer the contacts are to each other, the less regions away from them participate in the flow.

In the ballistic regime only discrete values of the Fermi momentum, corresponding to the classical trajectories between the entry and the exit, contribute to the conductance. This is attributed to the fact that for some, 'resonant', values of the Fermi momentum of the injected electrons, there exist families of trajectories connecting source and sink contacts in a direct way. Existence of those trajectories sharply increases conductivity. As a result, the conductance, as a function of the Fermi momentum, jitters and has a form of plateaus with oscillating peaks at particular values of the Fermi momentum. 

Combining these two behaviors would presumably lead to the disappearance of the plateaus and the suppression of oscillating peaks at a cross-over, which, in principle, could be observable experimentally. This set-up can serve as an exemplification of interaction enhanced conduction \cite{Guo2017}. 

\subsection{Slip Length extraction}
The circular junction set-up has one additional attractive feature: the curvature term and the microscopic slip-length in \ref{eq:slipeff} have opposite signs, which in the previously presented case of parallel flow leads to peculiar and probably unphysical behavior. So, we are led to expect distinctive effects when the microscopic slip is of order of the sample radius\footnote{Let us, however, stress, that the unphysical effects of the previous section originated in imposed simplifications, namely the assumption that the flow is strictly parallel even in a channel with (slightly) curved boundaries, so in the present case we do not expect such pathologies to occur.}. It turns out that one observable in which such an effect is visible is the boundary electrochemical potential profile, which can be computed from the stream function (see Appendix  C). This observable describes the local voltage drop along the boundary of the sample $\phi(\theta)$. In our case its measurement may allow us to directly access the slip length experimentally. \\
\begin{figure}[hbtp]
 \centering
 \includegraphics[width=0.47\textwidth]{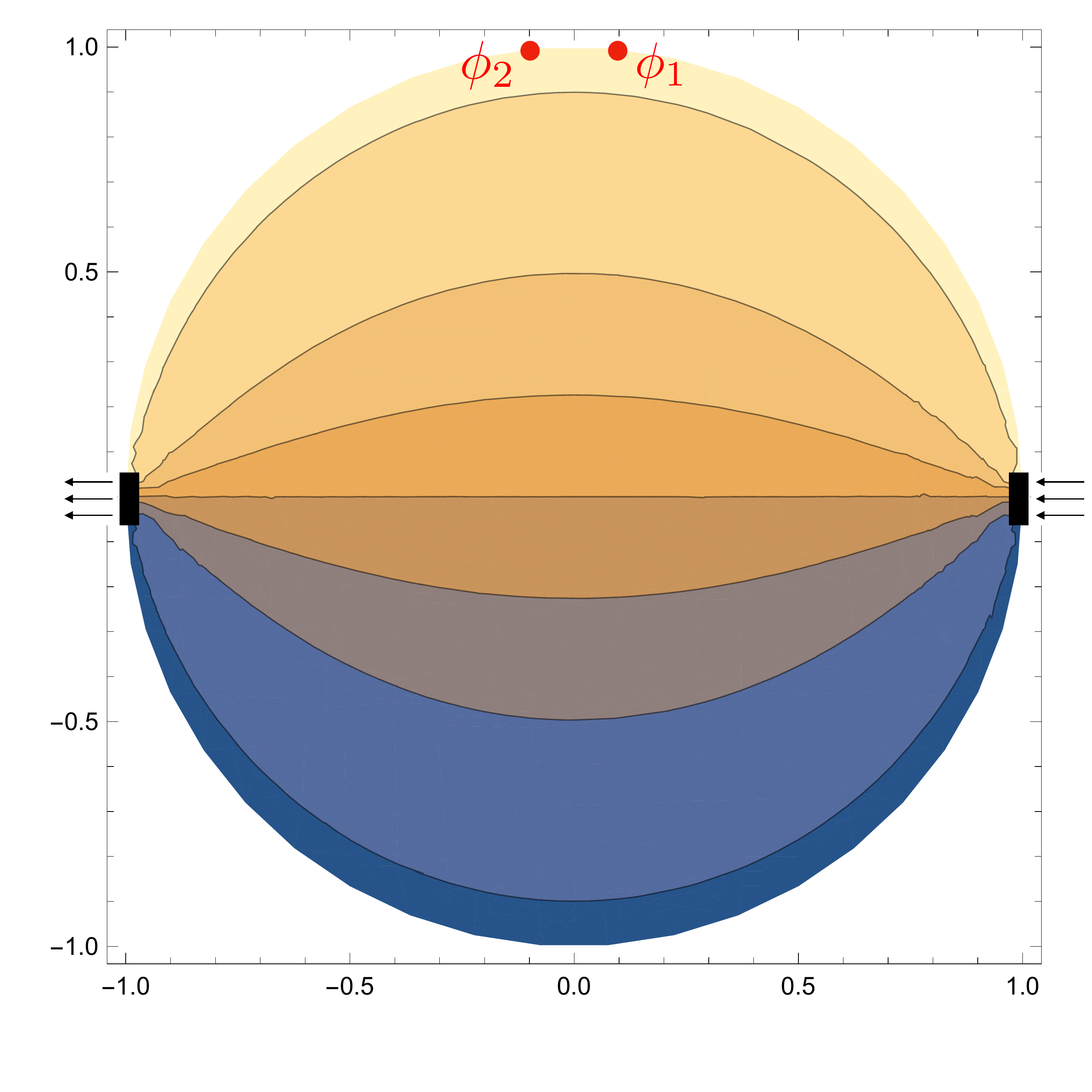}
  \caption{Experimental set-up proposed to extract the value of the microscopic slip-length. The current flows between a pair of contacts on opposite sides of circle diameter, and the electrochemical potential difference  $\phi_2-\phi_1$ is measured between two points}\label{pic:DiscMeasureSetUp}
 \end{figure}

 \begin{figure}[hbtp]
 \centering
\begin{tabular}{c}
(a)\\
\includegraphics[width=0.45\textwidth]{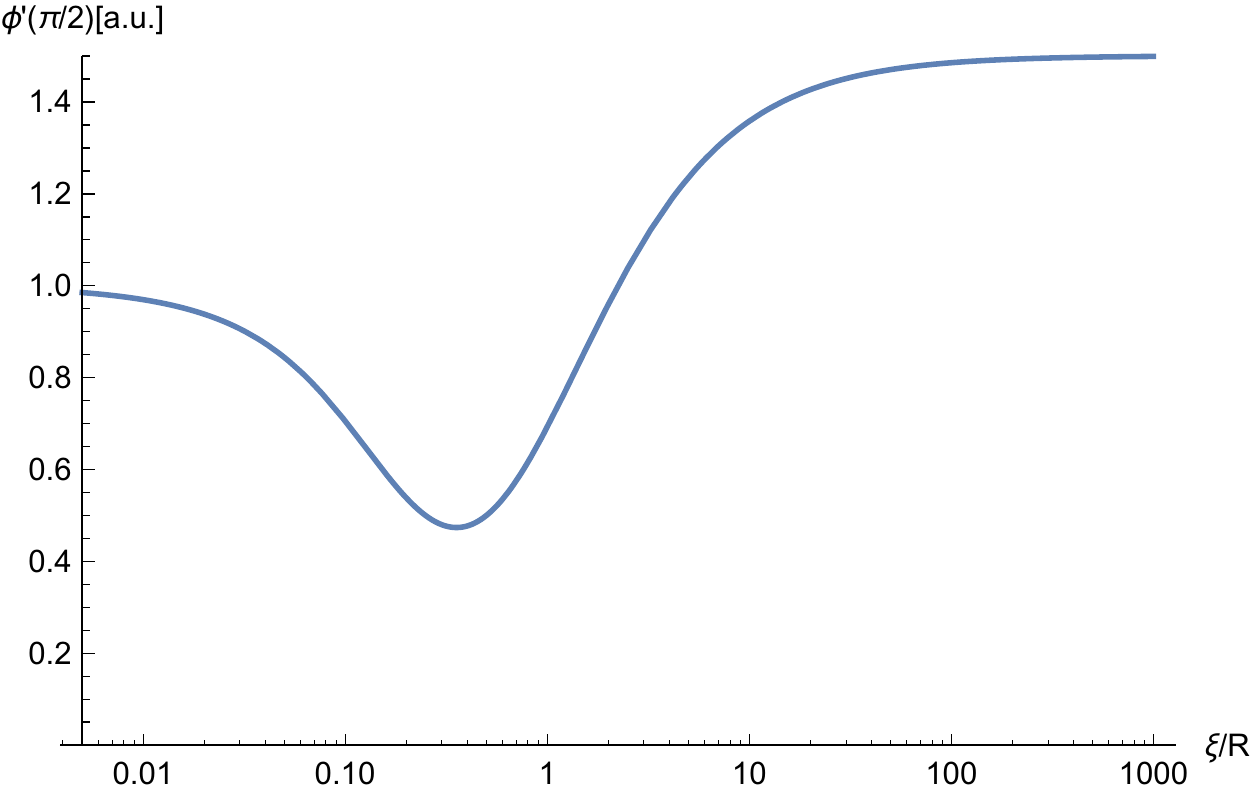} \\
%(b)\\
%\includegraphics[width=0.47\textwidth]{25micronrealtempNew.pdf} \\
(b)\\
\includegraphics[width=0.47\textwidth]{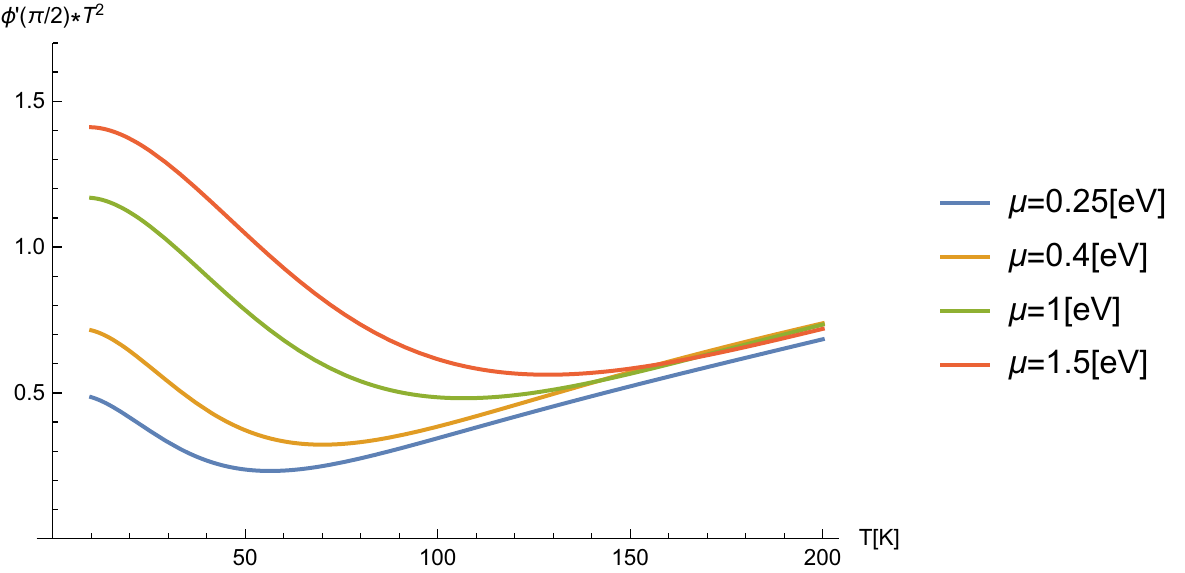} 
\end{tabular}
\caption{The proposed observable, boundary potential drop around $\pi/2$, is characterised by a strongly non-monotonic behaviour. \textbf{Top (a):} the observable as a function of a slip to radius ratio, keeping all the other parameters fixed. The plot is normalised to its no-slip value. 
%\textbf{Middle (b):} predicted values of the observable divided by $k$th power of temperature (to account for the thermal dependence of viscosity, $k=-2$ for doped case, $k=2$ at charge neutrality  \cite{narozhny2019magnetohydrodynamics}) as a function of temperature, assuming thermal slip dependences given by \cite{Kiselev2019} for doped ($\mu=0.4 eV$) graphene (blue) and charge neutral (renormalisation cut-off $\Lambda_r=1eV$) assuming nearly specular reflection (orange) or diffuse scattering (green) at the boundary. All the values of the parameters coincide with \cite{Kiselev2019} (microscopic edge roughness $h=h^\prime{}=250$\r{A}, electric permittivity $\varepsilon=5\varepsilon_0$, sample radius $2.5 \mu{}$m).
 \textbf{Bottom (b):} predicted values of the observable times temperature squared (to account for the thermal dependence of viscosity \cite{narozhny2019magnetohydrodynamics}) for doped graphene with various chemical potentials $\mu$. In the given range of dopings, the temperatures at which peaks occur lie in the hydrodynamic regime for graphene\cite{Bandurin2016}, which makes it feasible to measure the effect. The values of parameters used to generate the plot coincide with \cite{Kiselev2019}, all the plots are normalised by the room-temperature value of the observable ($T=293K$). The sample radius is $5 \mu$m%.
}\label{pic:BdrPotentialPeak}
 \end{figure}
The set-up we propose is the one presented on Fig. \ref{pic:Disc} with $\alpha=0,~\beta=\pi$. For computational convenience we also take contact sizes $\epsilon=\epsilon^{'}\rightarrow{}0$, so we inject current by point-like contacts. Remarkably, in this set up, the Fourier series can be summed up analytically and expressed in form of a rather complicated combination of hypergeometric functions for an arbitrary slip length.

A salient feature of the boundary potential profile is that the curves are not too distinct for no-stress and no-slip conditions, but have pronouncedly smaller derivative at an intermediate value of the slip length (numerically found to be $\xi/R \approx 0.36$). This is displayed in Fig. \ref{pic:BdrPotentialPeak} (a), which shows the value of  angular derivative of the potential precisely in the middle between two contacts as a function of microscopic slip. This quantity undergoes significant changes (around 50\%) upon changing microscopic slip from zero to infinity. So, we propose an experimental to measure the slip length would be to add to the set-up two measurement contacts located on the boundary around point $\theta=\pi/2$ (See fig. \ref{pic:DiscMeasureSetUp}). Then one would vary external conditions such as temperature or background chemical potential and observe the values of potential at the probe contacts $\phi_1,~\phi_2$. Such a set-up would effectively measure the angular derivative of the boundary electrochemical potential $\phi'(\pi/2)$.  

Panel (b) of Fig \ref{pic:BdrPotentialPeak} presents the predicted behavior of our observable as a function of temperature, assuming the slip length temperature dependence calculated in \cite{Kiselev2019}. We consider a case of doped graphene at different chemical potentials. It should be noted here, that the Stokes equations which we solve to obtain those results, are strictly speaking valid only in the Fermi liquid regime, i.e if the background chemical potential is much larger than the temperature. The main simplification in that regime is that thermal effects are suppressed, i.e. local temperature is no longer a relevant degree of freedom for the dynamics, and the electric charge current is proportional to the particle number current. That in turn limits the number of independent variables and equations, thus allowing for a simple description. 
The results discussed above, for doped samples, should be directly comparable to experiment\footnote{See appendix \ref{sec:potential} for a more detailed discussion of the employed approximations}.
It follows from the plots that the temperature dependence of the slip length can be measured using a series of such circular devices with different radii: for every radius $R$ there should be a temperature in which potential difference between the electrodes (divided or multiplied by $T^2$ for charge neutral and doped cases respectively, to get rid of thermal viscosity dependence) is maximal. Then, the slip length is approximately equal $0.36 R$ at this temperature. 

\section{Conclusions and discussion} In this paper we investigated the role of barriers on the channel walls as a tool to control boundary conditions in electronic flows. Through a numerical analysis we showed that in a system with large slip length, allowing in principle large velocities on the flow boundaries, we can nonetheless effectively realize a no-slip flow by introducing perpendicular barriers. The main motivation for this comes from the fact that large slip at the boundary hinders the hydrodynamic nature of the flow. As a result in graphene, where the slip velocity is believed to be large, the flow profile should depart from the parabolic Poiseuille profile. Engineering the no-slip boundary condition should facilitate the  experimental observation of viscous hydrodynamics.

In a more general context the message is that the properties of the surface affect the slip on the boundary. We have shown that the analysis of boundary conditions in the context of electronic fluid flow has to be modified to account for a mesoscopic boundary curvature and roughness as well as the boundary shape. As a result the effective slip velocity is not given purely by microscopics but rather a combination of the microscopic slip-length an sub-mesoscopic as well as mesoscopic curvature. We have constructed an explicit solution and the corresponding charge current under a general assumption of different effective slip-lengths at the boundaries of the two-dimensional flow.

 The contribution of the curvature is not limited to mesoscopic radii, in fact a sub-mesoscopically rough surface will also contribute to the effective slip. The consequences of this contribution have been ignored so far. A question that arises is when this is a legitimate thing to do. The answer is that we can ignore the microscopic roughness for boundaries which are characterized by a very diffuse scattering. One example may be the case of delafossite metals, where the samples are produced from flux-grown single crystals by focused ion beam etching. Ion beams produce boundaries which are diffuse. The microscopic slip-length contribution dominates over the geometric component and is well approximated by a no-slip condition. On the other hand bilayer
graphene devices are prepared using lithography and a subsequent etching processes. The boundary scattering is not efficient in dissipating momentum, which results in a large microscopic slip-length -- which may be comparable to the system size. In this case the mesoscopic, geometric contribution, being of the same order of magnitude, is important and should not be ignored.

Finally we studied a flow of electrons through a circular junction - a system with a fixed mesoscopic curvature. We found that, in analogy with the flow through a constriction, the hydrodynamic scenario provides for uniformly efficient transport. On the other hand, ballistic transport exhibits resonances attributed to special trajectories inside a junction corresponding to classical paths from the entrance to the exit of the junction. In hydrodynamics, because of frequent collisions between particles, the flow through the constriction is smooth.

To sum up, the set-ups with curved boundaries posses theoretical and practical advantages over their straight-bounded counterparts for studying viscous electronic flow. They may provide crisp signatures of viscous-to-ballistic crossover (resonant conductance in ballistic regime vs smooth in hydro) and they offer a possibility to directly experimentally access the microscopic slip length. As such, they call for more experimental attention then they have received so far.

\section*{Acknowledgments}We acknowledge useful conversations with Renato Dantas, Andrew Mackenzie, Philippa McGuinness, Nabhanila Nandi, Francisco Pe\~{n}a-Benitez, J\"{o}rg Schmalian, and Jonah Waissman. We would like to especially thank Egor Kiselev  who, apart from sharing his knowledge in many discussions also provided us with the formulae for the slip length temperature dependence that we used to produce the bottom panel of Fig. \ref{pic:BdrPotentialPeak}. This work was supported by the Deutsche Forschungsgemeinschaft via the Leibniz Programme as well as cluster excellence ct.qmat, EXC2147.

\appendix

\section{Solution in a channel with perpendicular obstacles}Following the procedure developed in \cite{Branicki2006}, we introduce the ansatz $\Psi(x,y,t)=\hat{\Psi}(x,y) \,e^{i\,\omega\,t}$, from which we obtain an equation for the spatial part of the stream function
\begin{align}
\label{biharmonic}
    \Delta^2\,\hat{\Psi}=\Lambda\,\Delta\,\hat{\Psi},
\end{align}
with $\Lambda=\frac{(i \omega+\gamma)(h/2)^2}{\eta}$. Note that for zero frequency $\Lambda=\Gamma$ is used in main text. This is the non-homogeneous biharmonic equation for the complex function $\hat{\Psi}$. We are interested in the real (physical) part of the complete stream function, which is given by
\begin{align}
    \text{Re}[\Psi]=\text{Re}[\hat{\Psi}]\cos(\omega\, t)-\text{Im}[\hat{\Psi}]\sin(\omega\,t).
\end{align}
We solve equation \eqref{biharmonic} for different values of the slip parameter and in different driving frequency regimes. In order to do so, we use the method of eigenfunction expansion and point match \cite{Wang1997}. Based on the symmetries that the geometry of the channel imposes on the stream function, we  define $\alpha=\frac{n\pi}{b}$ and $\beta=n\pi$ and propose an ansatz solution of the form
\begin{widetext}
\begin{align}
 \Psi(x,y)&=\frac{\sinh\left(\sqrt{\Lambda}\,y \right)-\sqrt{\Lambda}\cosh\left(\sqrt{\Lambda}\right)y}{\sinh\left(\sqrt{\Lambda} \right)-\sqrt{\Lambda}\cosh\left( \sqrt{\Lambda}\right)}+A_0\left[y-\frac{\sinh\left( \sqrt{\Lambda} y\right)}{\sinh\left( \sqrt{\Lambda}\right)} \right]\\
 &+\sum_{n=1}^{\infty}A_n\cos(\alpha x)P_n(y) +\sum_{n=1}^{\infty}\left[B_n\sin(\beta y)\,Q_n(x)+x\,C_n\sin(\beta y)\,T_n(x) \right]. \nonumber  
\end{align}
\end{widetext}
The first term corresponds to the solution for a flat channel \cite{Kiselev2019} and the second term introduces a possible correction due to the presence of the obstacles. The functions $P_n(y), Q_n(x)$ and $T_n(x)$ are given by
\begin{widetext}
\begin{align*}
    P_n(y)&=\left[e^{\sqrt{\alpha^2+\Lambda}\,  (y-1)}-e^{-\sqrt{\alpha^2+\Lambda}\,(y+1)}\right]-\frac{1-e^{-2\sqrt{\alpha^2+\Lambda}}}{1-e^{-2\alpha}} \left[e^{\alpha (y-1)}-e^{-\alpha(y+1)}\right].\\
    Q_n(x)&=e^{\sqrt{\beta^2+\Lambda}\,(x-b)}+e^{-\sqrt{\beta^2+\Lambda}\, (x+b)}, \hspace{1.5cm} T_n(x)=e^{-\beta (x+b)}-e^{\beta (x-b)}.
\end{align*}
\end{widetext}
We impose a fixed value for the stream function on the top and bottom walls and on the obstacles, $\psi(x,\pm h/2)=\pm 1$. Additionally we impose the slip boundary condition exactly on the top and bottom walls of the channel. For $x=b$ we choose $N$ equally-spaced points along this line and impose the boundary conditions there. Next, we truncate the series in the stream function up to $N$ terms, so that the problem reduces to solving a $(3N+1)\times (3N+1)$ resulting linear system. Once the coefficients $A_n$, $B_n$ and $C_n$ are obtained, the streamlines indicate the direction of the electron flow.

\section{Stability} \label{sec:stability} We want to analyze how the geometric contribution to the slip-length influences the stability of a flow. For negative curvature we expect that the flow will be stabilised as this case corresponds to a previously studied\cite{He2008} case of positive small slip-lengths. However, the case of positive curvatures, which can result in negative efective slip, has not been investigated and it may destabilise the Hagen-Poiseuille flow. To address this problem we employ linear stability analysis. In order to do so we find it convenient to use the stream function formulation of the Navier-Stokes equations and write down the Orr-Sommerfeld problem for linear perturbations \cite{orszag_1971,Spectral1,SchmidHenningson} \begin{equation}\label{eq:psi}
\Psi =\psi_0 +\psi.
\end{equation}
This is an eigenvalue equation describing the linear two-dimensional modes of disturbance to a viscous parallel flow. The perturbation has a wave-like structure
\begin{equation}\label{eq:psi1}
\psi = \text{exp}[i \alpha (y -c t)].
\end{equation}
\begin{figure}[hbtp]
\centering
\includegraphics[width=0.47\textwidth]{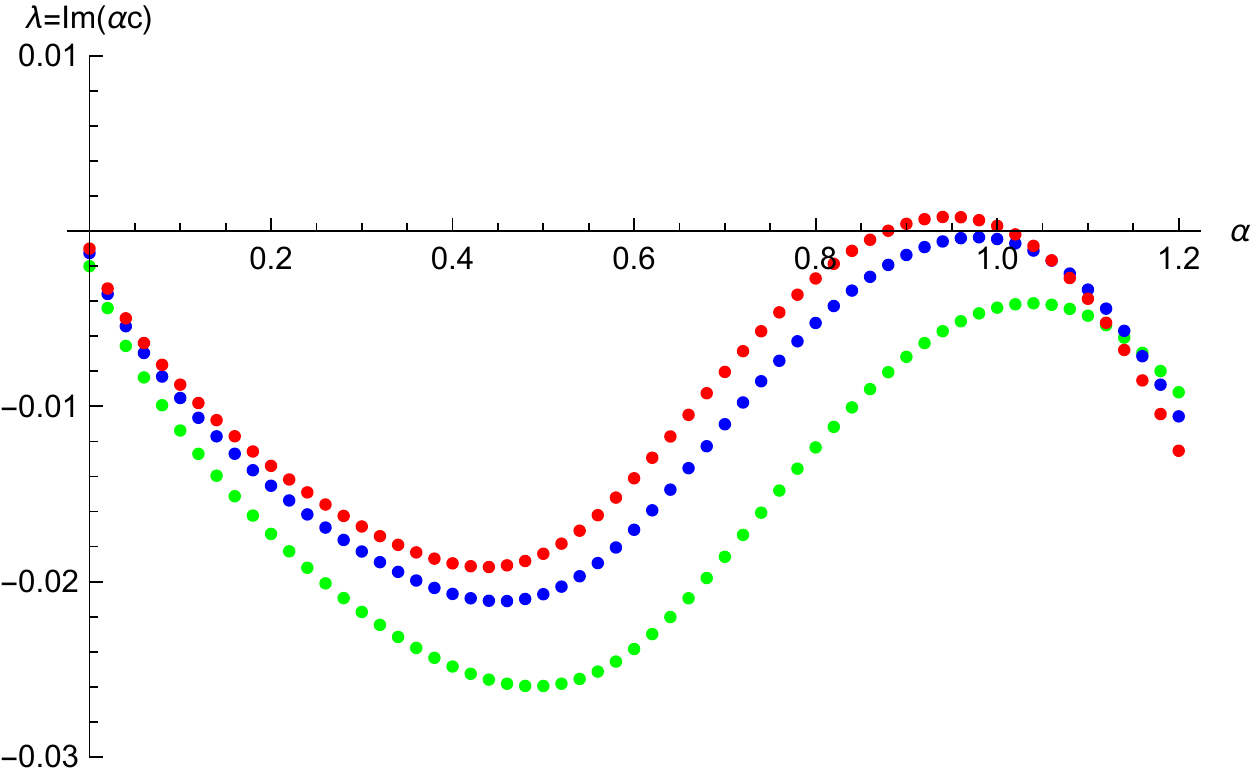} 
\caption{Dispersion curves of the Poiseuille flow for various Reynolds numbers. Green corresponds to $\text{Re}=5000$, blue to $\text{Re}=8000$, and red to $\text{Re}=10000$ with $\eta=1$, $\xi_1=\xi_2=-0.08$. We plot the dispersion relations for the modes which have the biggest imaginary part of frequency -- the dominant modes.}
\label{pic:m81000}
\end{figure}
The frequency $\omega=\alpha c$ is in general a complex number. Its imaginary part determines the stability of the flow -- if it is positive for some values of wave-number $\alpha$ or Reynolds number, Re, it indicates the existence of an unstable mode. This, once excited, will exponentially grow in time, ultimately destroying the solution. The flow is unstable if one or more eigenvalues $c$ have a positive imaginary part. Therefore, our goal is to check if such unstable modes are present, given the geometric contribution to the effective slip. The Orr-Sommerfeld equation for perturbations around the basic Hagen-Poiseuille flow reads
\begin{align}\label{eq:orr}
\frac{1}{i \alpha \text{Re}} \left( \frac{d^2}{dy^2}-\alpha^2\right)^2 \psi& -U \left( \frac{d^2}{dy^2}-\alpha^2\right)\psi+U'' \psi=\\ \nonumber
&-c  \left( \frac{d^2}{dy^2}\psi-\alpha^2\right)\psi,
\end{align}
where $U$ is the basic flow solution around which we perturb and $U''$ is the second $y$ derivative of $U$. We discretize the above equation using Chebyshev polynomials. The eigenfunctions are expanded in a basis defined on an interval $\{-1,1\}$ and we require that the equation \eqref{eq:orr} is satisfied at the Gauss-Lobatto collocation points $y_j=\text{cos}(\pi j/N) $. The final result is a generalized eigenvalue problem of the form
\begin{equation}
\mathbf{A} \, \mathbf{a} =c \,\mathbf{B}\,  \mathbf{a} .
\end{equation}
\begin{figure}[hbtp]
\centering
\includegraphics[width=0.47\textwidth]{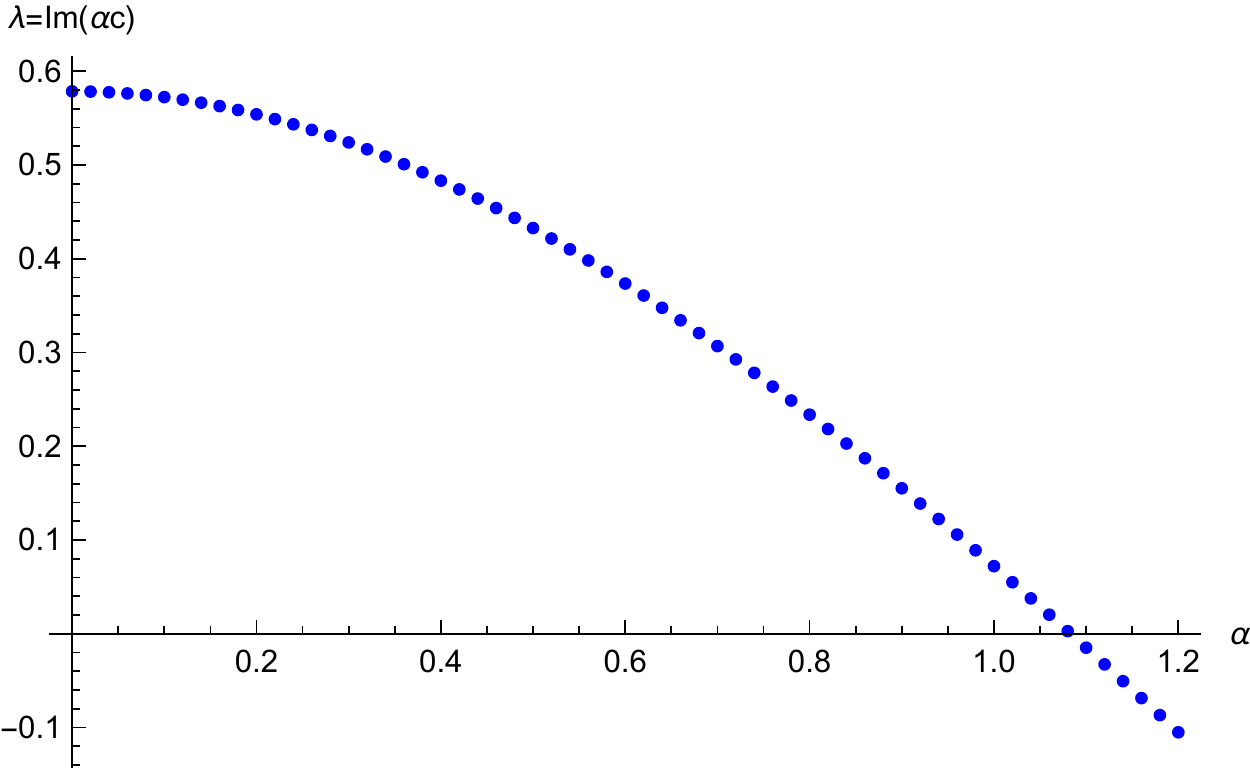} 
\caption{Dispersion curves of the Poiseuille flow for $\text{Re}=5$, with $\eta=1$, $\xi_1=\xi_2=-0.55$}
\label{pic:m551000}
\end{figure}
\begin{figure}[hbtp]
\centering
\includegraphics[width=0.47\textwidth]{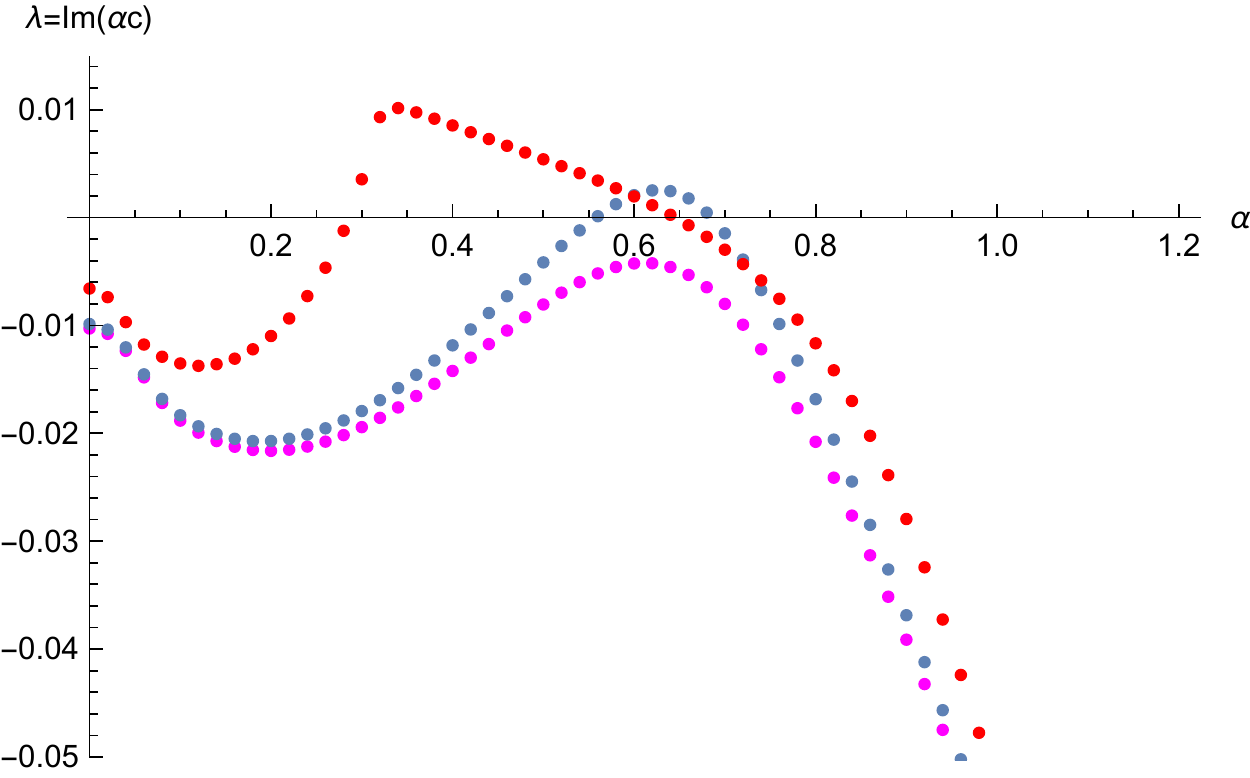} 
\caption{Dispersion curves of the Poiseuille flow for various Reynolds numbers. Magenta corresponds to $\text{Re}=960$, dark blue to $\text{Re}=1000$, and red to $\text{Re}=1500$ with $\eta=1$, $\xi_1=-0.1$, $\xi_2=0.1$. We plot the dispersion relation of the dominant modes.}
\label{pic:posneg01}
\end{figure}
Because the channel width is not the same as before, we need to construct the Poiseuille profile with the slip boundary conditions \eqref{eq:bc2} at $\{-1,1\}$. The solution reads
\begin{equation}
U(y)=\frac{1}{2\eta} \mathcal{\tilde{U}}(y) \frac{\partial p}{\partial x},
\end{equation}
where
\begin{equation}
\mathcal{\tilde{U}}(y) =1-y^2 +\left( \frac{2\xi_2-2\xi_1}{2+\xi_1+\xi_2} \right)+ \left( \frac{2\xi_2+2\xi_1+4\xi_1\xi_2}{2+\xi_1+\xi_2} \right).
\end{equation}
Our equation is fourth order in the derivatives, hence, we need to supplement two boundary conditions $\psi(1)=\psi(-1)=0$. Following the standard procedure we remove the first and the last two rows of the discretised equation to add boundary conditions and use $N=100$ collocation points. When both our slip-lengths are zero we recover the usual instability around $\alpha=1$ for $\text{Re}=10000$. If both slips-lengths are small, and positive, they stabilise the flow in accordance with previous studies \cite{He2008}. However, negative slip-lengths have a destabilising effect:
the instability is more pronounced the higher the absolute value of negative slip-lengths.
The same phenomenon is present if one boundary has a negative curvature, when even for quite low Reynolds numbers the flow becomes unstable.

\section{Inflow/outflow problem in a circular junction}
We consider the hydrodynamic flow inside a disc of unit radius. We construct a Fourier series solution from fundamental solutions of the biharmonic equation \cite{TikhonovSamarskii}
\begin{widetext}
\begin{equation}
\Psi (r,\theta) =a_0+b_0 r^2+ \sum _{n=1}^{\infty} (a_n r^n+b_n r^{n+2})\cos(n\theta)+ \sum _{n=1}^{\infty} (c_n r^n+d_n r^{n+2})\sin(n\theta).
\end{equation}
\end{widetext}
The coefficients $a_0$, $b_0$, $a_n$, $b_n$, $c_n$, $d_n$ are determined from the boundary conditions. In the problem of a flow through a disc, the boundary conditions are of the form
\begin{align}
    \Psi(1)&=f(\theta),\\
   \left. \frac{\partial^2 \Psi}{\partial r^2}\right| _{r=1}&=\left. \frac{\xi-1}{\xi}\frac{\partial \Psi}{\partial r} \right| _{r=1},\\
    \Psi(0)&=0.
\end{align}
We use the slip boundary condition \eqref{eq:GeneralBCflat} with the slip parameter depending on the curvature radius. Function $f(\theta)$ is fixed based on the number of inflow and outflow channels and the angular separation between them. For one inflow and one outflow contact corresponding to slit widths $\epsilon$ and $\epsilon'$ the function $f(\theta)$ is given by
\begin{equation}\label{eq:bc2_2}
    f(\theta)= 
\begin{cases}
    1+ \frac{\theta -\alpha}{\epsilon'}, &\alpha -\epsilon' < \theta<\alpha +\epsilon'\\\
    2, & \alpha +\epsilon' < \theta<\beta -\epsilon\\
   1+ \frac{\beta-\theta }{\epsilon},& \beta -\epsilon<\theta<\beta +\epsilon\\
    0,& \beta +\epsilon<\theta< 2\pi+\alpha-\epsilon'.
\end{cases}
\end{equation}
In the following we impose the condition that the inflow and outflow contacts have equal widths $\epsilon =\epsilon'$. Imposing the derivative condition on the boundary fixes
\begin{align}
    b_n&=a_n \frac{ n (1-n \xi  )}{(n+2) [(n+2) \xi -1]},\\
    d_n&=c_n\frac{ n (1-n \xi )}{(n+2) [(n+2) \xi -1]}.
\end{align}
The condition \eqref{eq:bc2_2}  allows one to determine $a_n$ and $c_n$ using the orthogonality condition \cite{Pinsky}
\begin{align}
    a_n&=-\frac{(n+2) [(n+2) \xi -1] \sin (n \epsilon ) [\sin (\alpha  n)-\sin (\beta  n)]}{\pi 
   n^2 \epsilon  [2 (n+1) \xi -1]},\\
    c_n&=\frac{(n+2) [(n+2) \xi -1] \sin (n \epsilon ) [\cos (\alpha  n)-\cos (\beta  n)]}{\pi 
   n^2 \epsilon  [2 (n+1) \xi -1]}.
\end{align}
Finally the condition at the origin fixes $a_0=0$. We note that	 for the most symmetric configurations with $\alpha=0,~\beta=\pi$ $a_n=b_n=0$. 
\section{Stream function formulation and electrochemical potential}\label{sec:potential}
In the hydrodynamic regime, the ensemble of electrons is  described in terms of the following equations: the Stokes equation
\begin{equation}
\partial_t u^i -\eta \Delta u^i + \gamma u^i = \frac{e}{m}\nabla{}^i\chi - \frac{1}{m}\nabla{}^i\delta \mu,\label{eq:Stokes1}
\end{equation}
and the continuity equation
\begin{equation}
\partial_j u^j=0.
\end{equation}
$\chi$ is the electric potential and $\delta \mu$ is the local variation of the chemical potential giving rise to effective pressure \cite{PellegrinoHall2017}, $m,~e$ are the mass and electric charge of a carrier respectively. In principle, the electric potential above should be not only the external driving potential but also should contain, even at the linearized level, a self-consistent term stemming from variations of a local carrier density \cite{Narozhny2017} that would make the equations non-local. However, since the electric potential enters in the equations in a special way, we can avoid this difficulty. 
First, we define the so-called electrochemical potential 
\begin{equation}
\phi = \chi -\frac{1}{e}\delta \mu,\label{eq:electrochemical}
\end{equation}
which combines both scalar functions in the Stokes equation into a single one. Real life experiments are usually sensitive to the electrochemical potential rather than electric voltage or chemical potential alone\cite{PellegrinoHall2017,Lucas2018}.
Taking that into account, the Stokes equation \ref{eq:Stokes1} is effectively incompressible, and the gradient of electrochemical potential can be decoupled from the system. To do this we introduce the stream function formulation of a viscous equation, in which one acts with antisymmetrized derivative on the Stokes equation to get rid of the gradient term. The electrochemical potential can be computed \emph{a posteriori} from the solution.\\
The procedure to do so is the following: we use the stream function definition  
\begin{equation}
u^i=\epsilon{}^{ij}\nabla_{j} \Psi{}
\end{equation}
in combination with Stokes equation sourced by the electrochemical potential \ref{eq:electrochemical}
\begin{equation}
\partial_t u^i -\eta \Delta u^i + \gamma u^i = \frac{e}{m}\nabla{}^i\phi.
\end{equation}
Then we solve the resulting equation
\begin{equation}
\nabla{}^i\phi=\frac{m}{e}\left(\partial_t \epsilon{}^{ij}\nabla_{j} \Psi{} -\eta \Delta \epsilon{}^{ij}\nabla_{j} \Psi{} + \gamma \epsilon{}^{ij}\nabla_{j} \Psi{} \right),
\end{equation}
which can be done by simple integration as now $\Psi$ is known solution. The function $\phi$ is the observed electrochemical potential.

\section{EPL slip deriviation}\label{sec:EPL}
Let us present the basic idea behind the EPL boundary condition. The discussion is based on \cite{panzer1992effects}. The origin of the modification of the microscopic slip length is the following: the force acting on surface of fluid element due to friction (drag) on a boundary is
\begin{equation}
dF^j=(\beta t_iu^i)t^j,
\end{equation}
with $\beta$ being the drag coefficient, and $t$ the normal vector, tangent to the boundary, provided that the boundary is at rest. This has to equal the viscous force acting on the fluid surface, given by
\begin{equation}
dF^k=dS\Pi^{k}_{i}n^i
\end{equation}
where $dS$ is the surface area of our element. Balancing those two forces yields
\begin{equation}
(\beta t_iu^i)t^j=dS\Pi^{k}_{i}n^i.
\end{equation}
Taking into account that 
$$
\Pi_{ij}=\tilde{\eta}\left(\partial_i u_j + \partial_j u_i\right)
$$ 
(where $\tilde{\eta}=\eta\rho$ is the \emph{dynamic viscosity}) and projecting the above on the transverse direction to get a scalar equation we obtain:
\begin{equation}
\left. u^i t_i \right|_B=\left.  \xi_0 t^i n^j \left(\partial_i u_j + \partial_j u_i\right)\right|_B,
\label{eq:BdrTensorilForm}
\end{equation}
 where the microscopic constants $\beta,~dS,~\eta$ are collected into a new one -- the (microscopic) slip length $\xi_0$, and the subscript $B$ means that the fields are evaluated on the boundary. $u$ is the velocity, $n$ is the \emph{inward} pointing normal, and $t$ is the unit tangent vector to surface.
Now, one can express $t^i$ and $n^i$ by the means of the parametric description of the boundary as a plane curve $x(s)$
\begin{equation}
t =\frac{1}{v}\frac{d}{ds} x,
\end{equation}
\begin{equation}
\frac{1}{R}n = \frac{1}{v} \frac{d}{ds} t,
\end{equation}
with  $v$ being the norm, i.e $v=|\frac{d}{ds} x|$, and $R (s)$ the local curvature radius\footnote{Here curvature $R$ is defined by the means of derivative of tangent vector in such a way to exhibit sign difference depending if domain is convex or concave. The given definition implies standard one, in the sense that $|R|=|\frac{d}{ds} t|^{-1}$} to obtain the result
\begin{equation}
u_T= \xi_0 \left[n^j\p^ju_T+u_T/R\right]
\end{equation}
with $u_T=u^i t_i$.  The latter expression can be re-arranged into the familiar form \ref{eq:GeneralBCflat} upon defining
\begin{equation}\label{eq:slipeff2}
\xi_\text{eff}= \left( \frac{1}{\xi_0}-\frac{1}{R}\right) ^{-1}.
\end{equation}
We emphasize that the conventions for curvature radius are such that for convex domains $R>0$ and for concave ones $R<0$.

The modification of the slip can also be thought of as coming directly from the tensorial form of boundary condition \ref{eq:BdrTensorilForm} in any curvilinear coordinate system. If one wants to write it in arbitrary coordinates, it takes form
\begin{equation}
\left. \left(u^i t_i-\frac{\xi_0}{\eta}\Pi_{ij}t^in^j \right)\right|_B = 0,
\end{equation}
where the shear stress tensor $\Pi$  is now given in terms of velocity
\begin{equation}
\Pi_{ij}=2\eta\nabla_{(i}u_{j)},
\end{equation}
with $\nabla$ being the covariant derivative\footnote{Mind that although the metric is flat Cartesian one, the connection coefficients in arbitrary coordinates may be nonzero.}. Now, let us use polar coordinates $\{r,\theta\}$ as an example. Let the domain boundary be a circle of radius $R$. The normal and tangent vectors are then basis vectors $\hat{r}$ and $\hat{\theta}$ respectively so the boundary condition turns into
\begin{equation}
u_\theta-\xi_0\Pi_{r \theta}=0
\label{eq:force}
\end{equation}
The viscous stress tensor evaluated on a boundary of curved domain is given by
\begin{equation}\label{eq:stress}
\Pi_{r \theta}=\eta \left( \frac{\partial u_\theta}{\partial r} -\frac{u_\theta}{R} \right).
\end{equation}
Plugging this expression back to \eqref{eq:force} allows us to define an effective slip-length \eqref{eq:slipeff}. So we see how the curvature enters the boundary condition and modifies slip.
\bibliographystyle{apsrev4-1}

\bibliography{GeomSlip}

\end{document}